\documentclass{aa}  

\usepackage{graphicx}
\usepackage{placeins}
\usepackage{xcolor}
\usepackage{mathtools}
\usepackage{bbm}
\usepackage{placeins}
\usepackage{url}
\usepackage{amsmath}
\usepackage{multirow}
\usepackage{txfonts}
\newcommand{\D}[1]{{\rm d} #1}

\newcommand{\m}{$^{\rm m}$}

\newcommand{\gaia}{{\it Gaia}}

\begin{document}

   \title{Analysing spectral lines in \textit{Gaia} low-resolution spectra}

   \author{M. Weiler\inst{1,2,3}, J.~M. Carrasco\inst{1,2,3}, C. Fabricius\inst{1,2,3}, and C. Jordi\inst{1,2,3}
             }

\authorrunning{M. Weiler et al.}

   \institute{Departament de F{\'i}sica Qu{\`a}ntica i Astrof{\'i}sica (FQA), Universitat de Barcelona (UB), c. Mart{\'i} i Franqu{\`e}s, 1, 08028 Barcelona, Spain
   \and
   Institut de Ci{\`e}ncies del Cosmos (ICCUB), Universitat de Barcelona (UB), c. Mart{\'i} i Franqu{\`e}s, 1, 08028 Barcelona, Spain
   \and
   Institut d'Estudis Espacials de Catalunya (IEEC), c. Gran Capit{\`a}, 2-4, 08034 Barcelona, Spain\\
              \email{mweiler@fqa.ub.edu}
             }

   \date{Received 18 August 2022; accepted 12 November 2022}

  \abstract
   {With its third data release, European Space Agency's \gaia~mission publishes for the first time low resolution spectra for a large number of celestial objects. These spectra however differ in their nature from typical spectroscopic data. They do not consist of wavelength samples with associated flux values, but are represented by a linear combination of Hermite functions.} 
   {We derive an approach to the study of spectral lines that is robust and efficient for spectra that are represented as a linear combination of Hermite functions.}
   {For this purpose, we combine established computational methods for orthogonal polynomials with the peculiar mathematical properties of Hermite functions and basic properties of the \gaia~spectrophotometers. Particular use is made of the simple computation of derivatives of linear combinations of Hermite functions and their roots.}
   {A simple and efficient computational method for deriving the position in wavelength, the statistical significance, and the line strengths is presented for spectra represented by a linear combination of Hermite functions. The derived method is fast and robust enough to be applied to large numbers of \gaia~spectra without high performance computing resources or human interaction. Example applications hydrogen Balmer lines, He I lines, and a broad interstellar band in \gaia~DR3 low resolution spectra are presented.}
{}
 

   \maketitle
%

\section{Introduction \label{sec:introduction}}

The European Space Agency mission \gaia~\citep{Gaia2016a} repeatedly scans the entire sky and observes point-like or nearly point-like objects at magnitudes from approximately $G=3$\m~to $G=20.7$\m. These observations include aperture-free low resolution spectrophotometry covering a wavelength range from approximately 330~nm to 1050~nm, with a spectral resolution of about 30 to 100 \citep{Carrasco2021}. The third \gaia~Data release (\gaia~DR3, \citealt{Vallenari2022}) includes these spectrophotometric observations for about 219 million astronomical objects \citep{DeAngeli2022}. The purpose of these spectra is supporting the calibration of the astrometric observations of \gaia, and to characterise the observed astronomical objects \citep{Gaia2016a}. Spectroscopy at very low resolution is naturally not well suited for the study of spectral lines, as the contrast between the line and the underlying continuum is poor, the shape of the lines is not or only partially resolved, and blending of different lines might be strong. However, the exceptional properties of the \gaia~spectrophotometry, such as the all-sky coverage from a stable space environment, its large degree of completeness, down to $G =17.65$\m~in \gaia~DR3, and the wide wavelength range covered, motivates efforts for analysing spectral lines in \gaia's low resolution spectra as well. Eventually, \gaia's final data release will comprise hundreds of low resolution epoch spectra collected over several years, and completeness to about $G=20.7$\m.\par
As a consequence of the self-calibration strategy for \gaia~spectra, they are represented by a linear combination of basis functions \citep{Carrasco2021}. This distinguishes \gaia~low resolution spectra from typical spectra, which consist of individual flux values on a grid of wavelength points. The implications of this difference in spectral representation have to be taken into account when answering questions such as, are spectral lines present in the \gaia~low resolution spectrum for a given astronomical object and at what level of statistical significance? Where are the lines located in wavelength, and what is the associated uncertainty? What is the line strength? Is the line broad and can we learn from its shape? In the following we are working out answers to these questions.\par
The structure of this work is as follows. In Sect. \ref{sec:basisFunctionRepresentation} we describe in detail the differences between typical spectra and spectra represented by a linear combination of basis functions, as is the case for \gaia. The unconventional properties of spectra of the latter kind are worked out there. Section~\ref{sec:problem} summarises the particular properties of \gaia~DR3 low resolution spectra. In Sect. \ref{sec:mathematicalProperties} we summarise the relevant mathematical properties of Hermite functions and derive mathematical relations required in this work. In Sect. \ref{sec:spectralFormation} we address relevant aspects of the formation of \gaia~spectra and in Sect. \ref{sec:inversion} we discuss the problem of inverting spectra. In the following sections, we discuss spectral lines as local extrema (Sect.~\ref{sec:localExtrema}), discuss the cases of narrow (Sect.~\ref{sec:narrowLines}) and broad (Sect.~\ref{sec:broadLines}) lines, and the use of higher derivatives (Sect.~\ref{sec:higherOrder}). Finally, we provide some example applications to \gaia~DR3 spectra in Sect.~\ref{sec:Examples} and close this work with a summary and discussion in Sect.~\ref{sec:summary}.

\section{Representing spectra with basis functions \label{sec:basisFunctionRepresentation}}

\begin{figure}
   \centering
   \includegraphics[width=0.49\textwidth]{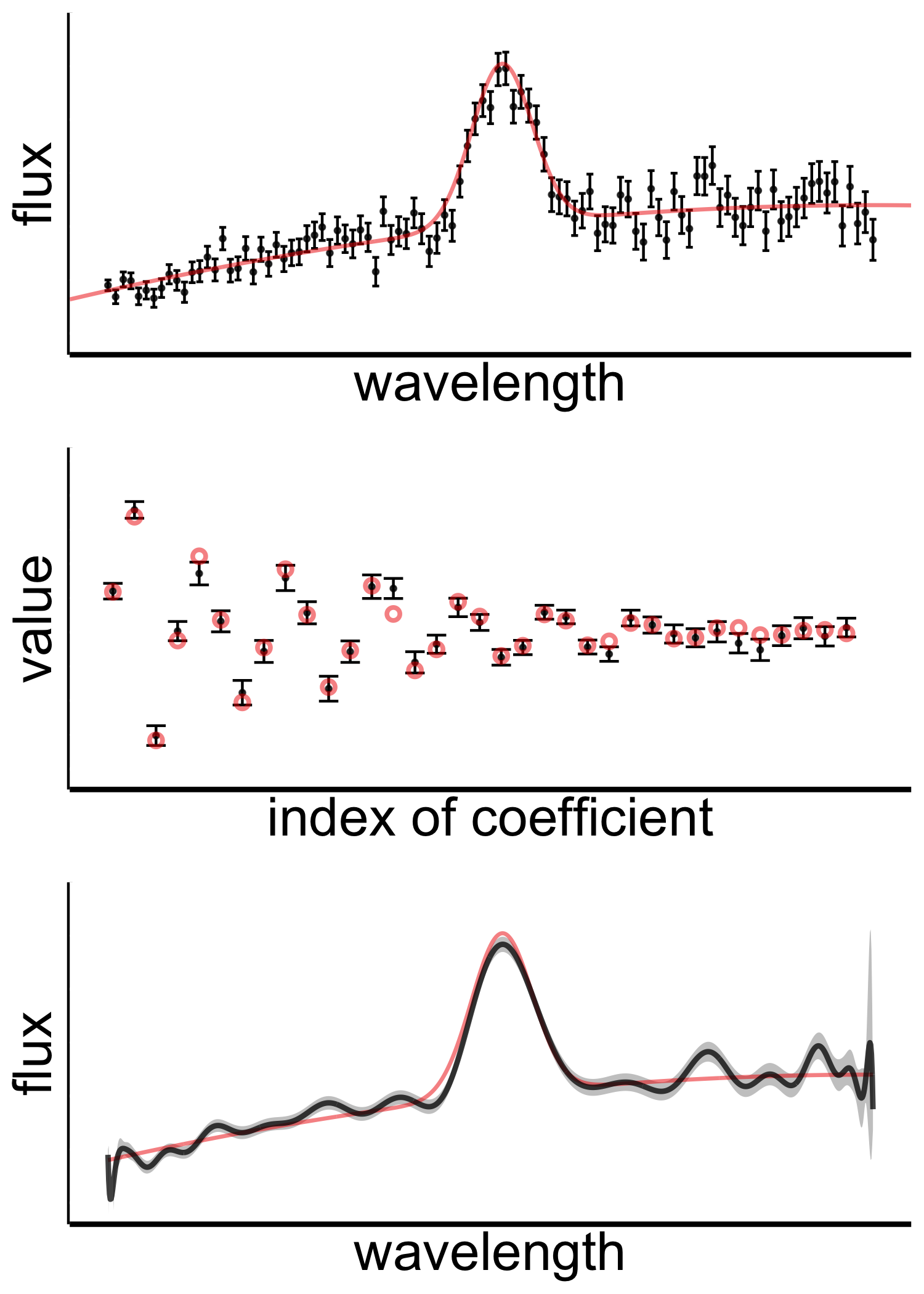}
   \caption{Sketch illustrating the difference between a typical spectrum and a spectrum represented by a linear combination of basis functions. Top panel: A typical spectrum, consisting of sampled flux values on a wavelength grid (black symbols) derived from the spectral photon distribution (red line). Central panel: Coefficients representing the same data set by a linear combination of basis functions. Black symbols show the noisy representation, red dots the true coefficients. Bottom panel: The linear combination of basis functions (black line) and the 1-$\sigma$ error interval (grey shaded region) as a function of wavelength.}
              \label{fig:sketch}
    \end{figure}

Typical astronomical spectra consist of flux measurements at different wavelength points, or bins. The flux measurements usually have low correlations, and are significantly correlated only over very short distances in wavelength, if the line spread function (LSF, the point spread function integrated along the direction perpendicular to the dispersion direction) is oversampled. This is illustrated in the sketch in the top panel of Fig.~\ref{fig:sketch}. The true spectral photon distribution (SPD, red line) is sampled at certain wavelength points, with uncorrelated normally distributed random noise (black symbols) in this example. A spectral line, an emission line in the sketch, manifests by flux points deviating from an assumed continuum, and the significance of the line can, in principle, be derived from how many standard deviations points deviate from the continuum.\par
If the same data set is instead represented by a linear combination of basis functions, then the coefficients of the development tend to have low correlations over small distances in coefficient indices. This is shown as a sketch in the central panel of Fig.~\ref{fig:sketch}, for the same set of data points. From these coefficients, the spectrum can be constructed by computing the actual linear combination on a continuous wavelength axis. This is shown in the bottom panel of Fig.~\ref{fig:sketch}. Here, noise manifests not as scatter of points around a mean, as is the case for the typical spectrum in the top panel, but in continuous, wavy structures. The shaded region in the bottom panel of Fig.~\ref{fig:sketch} gives the formal 1-$\sigma$ error interval around the mean spectrum. It is not trivial to judge visually, which wavy pattern is noise, and which actually is a spectral line. In this work, we derive a formalism to detect lines, ideally from the coefficients describing the spectrum directly, as well as a statistics to quantify the significance of wavy structures in \gaia's low resolution spectra. 

\section{\textit{Gaia} low resolution spectra \label{sec:problem}}

We summarise in this section the properties of \gaia~low resolution spectra as far as they are relevant in the context of this work. This summary is necessarily brief and incomplete and it is a compilation of information provided by \cite{Carrasco2021} for the principles of the calibration of inner-instrumental effects, by \cite{DeAngeli2022} for the actual processing and validation of \gaia~DR3 data, and by \cite{Montegriffo2022} for the absolute flux and wavelength calibration. The reader is referred to these works for more in depth discussions.\par
The \gaia~low resolution spectrophotometer consists of two parts, producing independent spectra for each observed celestial object on the wavelength intervals of approximately 330 nm to 680 nm and 640 nm to 1050 nm, respectively. The spectra on the short wavelength range are referred to as the ''Blue Photometer'', or, BP spectra, and the spectra for the longer wavelength range as the ''Red Photometer'', or, RP spectra. We use the term XP spectra to refer to both types of \gaia~spectra. From observed XP spectra of a particular source, transiting \gaia's fields of view repeatedly, ''mean'' BP and RP spectra are constructed. In this process, referred to as the ''internal calibration'', inner-instrumental effects, e.g. variations of response or line spread functions with time and position in the focal plane, are suppressed. The mean BP and RP spectra for an astronomical object are represented by a linear combination of 55 Hermite functions in BP and RP, respectively, in \gaia~DR3. The coefficients of these representations, together with the corresponding errors and correlations, are part of \gaia~DR3. The argument of the Hermite functions is a linear function of pseudo-wavelength, $u$, which is a coordinate along the wavelength dimension in the focal plane common to all BP and RP spectra, respectively, and which corresponds to a mean CCD pixel scale. The dispersion function links the pseudo-wavelength to the actual wavelength, $\lambda$. The value of the linear combination of Hermite functions is the flux of the source, in photo-electrons per second, per unit of pseudo-wavelength and within the \gaia~aperture of 0.7872~m$^2$. The coefficients of the linear combination refer not to Hermite functions directly, but to an orthogonal transformation of Hermite functions. The orthogonal transformations are different for BP and RP, and the transformation matrices are available with \gaia~DR3.\par
For the 55 orthogonal transformed Hermite functions, corresponding basis functions in absolute flux and wavelength units are provided with \gaia~DR3. The same coefficients that represent the internally calibrated BP or RP spectrum, also represent the spectral energy distribution when using the corresponding external basis functions. The transformation of internal spectra to external spectra however is intrinsically an ill-conditioned problem \citep{Weiler2020} and thus sensitive to systematic errors introduced to what essentially is a deconvolution process. For the analysis of spectral lines, the use of internally calibrated spectra is therefore preferable. Furthermore, the Hermite functions used in the representation of internally calibrated BP and RP spectra provide very useful mathematical properties that we can exploit in the analysis of spectral lines.

\begin{table}
\renewcommand\arraystretch{1.2}
\caption{List of notations used throughout this work.}
\begin{tabular}{l l}\hline
Symbol & Meaning \\ \hline
$\alpha$ & integral over $s_L(\lambda)$\\
$\bf B$ & nonstandard companion matrix to $\bf c$\\
$\bf c$ & coefficient vector of observational spectrum\\
${\bf D}^k$ & matrix transforming to $k-$ derivative\\
$\phi(x)$ & Linear combination of Hermite functions\\
$\varphi_n(x)$ & $n-$th Hermite function \\
${\bf \Phi}(x)$ & vector of values of $\varphi_n(x),\,n=0,\ldots,N-1$\\
$f(u)$ & observational spectrum\\
$f_c(u)$ & continuum contribution to $f(u)$\\
$f_l(u)$ & line contribution to $f(u)$\\
$g^{(k)}(x)$ & $k-$th derivative of a function $g(x)$\\
$\bf i$ & vector with integrals of Hermite functions\\
$\lambda$ & wavelength \\
$\lambda_L$ & wavelength of a narrow line\\
$L(u,\lambda)$, $L(u,u^\prime)$ & (pseudo-)line spread function\\
$N$ & number of coefficients in $\bf c$\\
$R(\lambda)$, $R(u^\prime)$ & response function \\
$s(\lambda)$, $s(u^\prime)$ & spectral photon distribution\\
$s_c(\lambda)$, $s_c(u^\prime)$ & continuum contribution to $s(\lambda)$, $s(u^\prime)$\\
$s_l(\lambda)$, $s_l(u^\prime)$ & line contribution to $s(\lambda)$, $s(u^\prime)$\\
$\Sigma_x$ & covariance matrix for vector $\bf x$\\
$u$ & pseudo-wavelength \\
$u^\prime$ & pseudo-wavelength corresponding to $\lambda$\\
$u_E$ & pseudo-wavelength for local extremum \\
$u_L$ & pseudo-wavelength for spectral line \\
$W$ & equivalent width of a spectral line\\
\end{tabular}
\end{table}

\section{Some mathematical properties of linear combinations of Hermite functions \label{sec:mathematicalProperties}}

In this section we compile and derive the mathematical properties and relationships involving Hermite functions that we need in the context of this work. We separate the mathematical aspects in this section from instrumental and astrophysical aspects discussed in the following sections. By doing so, the structure of this work becomes simpler, as we do not have to interrupt physical and astrophysical discussions with introducing the special mathematical properties of the Hermite functions later in the work. Table~1 provides an overview of the nomenclature that is used throughout this work for easy reference.

\subsection{Hermite functions and linear combinations thereof}

Hermite functions appear in a number of mathematical and scientific contexts. In physics, they own their relevance widely to being the eigenfunctions of the quantum mechanical harmonic oscillator. A detailed discussion of Hermite functions and their properties may therefore be found in most basic text books on quantum mechanics (e.g. \citealt{Magnasco2013}). The $n-$th Hermite function we denote with $\varphi_n(x)$, with $n=0,1,2,\ldots$. The index $n$ we refer to as the ''order'' of the Hermite function. In the context of this work we need two basic recurrence relations for Hermite functions $\varphi_n(x)$, $n=0,1,2,\ldots$. The first recurrence relation is
\begin{equation}
x\, \varphi_n(x) = \sqrt{\frac{n}{2}}\, \varphi_{n-1}(x) + \sqrt{\frac{n+1}{2}} \, \varphi_{n+1}(x) \quad , \label{eq:rec1}
\end{equation}
together with
\begin{eqnarray}
\varphi_0(x) &=& \pi^{-1/4} \, {\rm e}^{-\frac{1}{2}\, x^2} \nonumber \quad , \\ 
\varphi_1(x) &=& \pi^{-1/4} \sqrt{2} \, x \, {\rm e}^{-\frac{1}{2}\, x^2} \quad .  \nonumber 
\end{eqnarray}
From this relation between Hermite functions of different order $n$, and the first two functions given explicitly, any Hermite function can easily be computed for any argument $x$.\par
The second recurrence relation connects the first derivative of an Hermite function with Hermite functions. In this work, we denote the $k-$th derivative of a function with the superscript ''$(k)$''. For Hermite functions, we have the relation
\begin{equation}
\varphi_n^{(1)}(x) = \sqrt{\frac{n}{2}}\, \varphi_{n-1}(x) - \sqrt{\frac{n+1}{2}}\, \varphi_{n+1}(x) \quad . \label{eq:rec2}
\end{equation}
The Hermite functions are a set of orthonormal functions on $\mathbb R$, i.e. the integrals over the product of two Hermite functions over the entire real axis obeys the orthonormality relation
\begin{equation}
\int \limits_{-\infty}^{\infty} \varphi_i(x) \cdot \varphi_j(x) \, \D{x} = \delta_{ij} \quad , \label{eq:orthonormality}
\end{equation}
with $\delta_{ij}$ the Kronecker delta. We furthermore consider linear combinations of Hermite functions, e.g. functions of the form
\begin{equation}
\phi(x) = \sum\limits_{n=0}^N\, c_n \, \varphi_n(x) \quad , \label{eq:linearCombination}
\end{equation}
with the $c_n$ the coefficients of the linear combination. We may write the $N+1$ coefficients as an $N+1$ element vector $\bf c$. With the orthonormality expressed in Eq.~(\ref{eq:orthonormality}), the coefficients $c_n$ in Eq.~(\ref{eq:linearCombination}) become
\begin{equation}
c_n = \int\limits_{-\infty}^{\infty} \phi(x) \cdot \varphi_n(x) \, \D{x} \quad . \label{eq:coefficients}
\end{equation}

\subsection{Integrals over linear combinations of Hermite functions}
Equation (\ref{eq:orthonormality}) makes computations of integrals over the real axis over products of Hermite functions trivial. Using the relationships of Eq. (\ref{eq:rec2}) also makes the computation of integrals over finite intervals $[a,b]$ on the real axis simple. We get
\begin{equation}
\int\limits_a^b \varphi^{(1)}_n(x)\, \D{x} = \sqrt{\frac{n}{2}}\, \int\limits_a^b \varphi_{n-1}(x) \, \D{x}- \sqrt{\frac{n+1}{2}}\, \int\limits_a^b \varphi_{n+1}(x) \, \D{x} \quad ,
\end{equation}
and therefore
\begin{equation}
\int\limits_a^b \varphi_{n+2}(x)\, \D{x} = \sqrt{\frac{n+1}{n+2}}\, \int\limits_a^b \varphi_n(x)\, \D{x} + \sqrt{\frac{2}{n+1}}\, \left[\varphi_{n+1}(a) - \varphi_{n+1}(b) \right] \quad .
\end{equation}
Together with the indefinite integrals for the first two Hermite functions \citep{Brychkov1989},
\begin{eqnarray}
\int \varphi_0(x)\, \D{x} & = & \frac{1}{\sqrt{2}}\, \pi^{-\frac{1}{4}} \, \text{erf}\left(\frac{x}{\sqrt{2}}\right) \\
\int \varphi_1(x)\, \D{x} & = & -\sqrt{2}\, \pi^{-\frac{1}{4}} \, {\rm e}^{-\frac{x^2}{2}} \quad ,
\end{eqnarray}
the integrals over any interval on the real axis can be computed iteratively for all orders. In particular for integrations of the entire real axis, that is $a \equiv -\infty$ and $b \equiv \infty$, we obtain
\begin{equation}
\int \limits_{-\infty}^{\infty} \varphi_{2n+1}(x) \, \D{x} = 0
\end{equation}
and the integrals over the even order Hermite functions can be computed from 
\begin{eqnarray}
\int \limits_{-\infty}^{\infty} \varphi_{0}(x) \, \D{x} & = &  \pi^{\frac{1}{4}} \sqrt{2} \quad , \\
\int \limits_{-\infty}^{\infty} \varphi_{2(n+1)}(x) \, \D{x} & = & \sqrt{\frac{n+\frac{1}{2}}{n+1}} \, \int \limits_{-\infty}^{\infty} \varphi_{2n}(x) \, \D{x} \quad .
\end{eqnarray}
When arranging the integrals over the interval $[a,b]$ for the first $N$ Hermite functions as an $N$-element vector $\bf i$, then the integral over the linear combination of Hermite functions $\phi(x)$ from Eq. (\ref{eq:linearCombination}) over the real axis is
\begin{equation}
\int \limits_{-\infty}^{\infty} \phi(x)\, \D{x} = {\bf i}^{\mathsf T} \, {\bf c} \quad . \label{eq:integral}
\end{equation}
Since the vector {\bf i} only depends on the Hermite functions, but not on the particular linear combination thereof, the integral over any linear combination of Hermite functions can eventually be computed by the cost of one vector multiplication.

\subsection{Derivatives of linear combinations of Hermite functions}
The recurrence relation given by Eq.~(\ref{eq:rec2}) has immediate important consequences, as the derivative of an Hermite function is a linear combination of two other Hermite functions, with orders increased and reduced by one. And since differentiation is a linear operation, it follows that the derivative of any linear combination of Hermite functions is another linear combination of Hermite functions. Thus, the $k-$th derivative of a linear combination of Hermite functions with maximal order $N$ is a linear combination of Hermite functions of maximal order $N+k$.\par
For linear combinations of Hermite functions, the differentiation is therefore conveniently expressed by a matrix operation. We define the $(N+1) \times N$ matrix ${\bf D}^{1}$ as having zero entries except for the two sub-diagonals, whose entries are
\begin{equation}
{\bf D}^{1}_{i,i+1}  =  \sqrt{\frac{i}{2}} = -{\bf D}^{1}_{i+1,i} \quad .
\end{equation}
Then the derivative of a linear combination of the first $N$ Hermite functions, with coefficient vector $\bf c$, is the linear combination of the first $N+1$ Hermite functions, with the coefficient vector ${\bf c}^{(1)}$ given by ${\bf c}^{(1)} = {\bf D}^{1} \, {\bf c}$.\par
We may generalise the notation to the $k-$th derivative, if we expand the size of the matrix ${\bf D}^{1}$ to $(N+k) \times (N+k)$, and assume zero-padding for $\bf c$, i.e. assuming $\bf c$ being of length $N+k$, with all entries $N < n \le N+k$ being zero. Thus we have
\begin{equation}
{\bf c}^{(k)} = {\bf D}^{k} \, {\bf c} \quad , \label{eq:derivative}
\end{equation}
and the $k-$th derivative matrix ${\bf D}^{k}$ we obtain from $k$ multiplications of ${\bf D}^{1}$, i.e. ${\bf D}^{k}$ is the $k-$th power of ${\bf D}^1$. For ${\bf D}^{2} = {\bf D}^{1}\, {\bf D}^{1}$ we obtain a matrix whose elements are
\begin{equation}
{\bf D}^{2}_{i,j}  =
\begin{cases}
	\; \frac{1}{2} - i & \text{ for } i = j\\
	\; \frac{1}{2} \sqrt{i(i+1)} & \text{ for } i-j = -2  \\
 	\; \frac{1}{2} \sqrt{j(j+1)} & \text{ for } i-j = 2   \\
	\; 0 & \text{else} \quad .
\end{cases}
\end{equation}
For illustration, the first $5 \times 5$ elements of ${\bf D}^{1}$ and ${\bf D}^{2}$ are
\begin{equation}
{\bf D}^{1} =
\begin{pmatrix}
0 & \sqrt{\frac{1}{2}} & 0 & 0 & 0 \\
-\sqrt{\frac{1}{2}} & 0 & \sqrt{\frac{2}{2}} & 0 &  0 \\
0 & -\sqrt{\frac{2}{2}} & 0 &  \sqrt{\frac{3}{2}} & 0  \\
0 & 0 & -\sqrt{\frac{3}{2}} & 0 &\sqrt{\frac{4}{2}}  \\
0 & 0 & 0 & -\sqrt{\frac{4}{2}} & 0
\end{pmatrix}
\end{equation}
and
\begin{equation}
{\bf D}^{2} =
\begin{pmatrix}
-\frac{1}{2} & 0 & \sqrt{\frac{1}{2}} & 0 & 0 \\
0 & -\frac{3}{2} & 0 & \sqrt{\frac{3}{2}} &  0 \\
\sqrt{\frac{1}{2}} & 0 & -\frac{5}{2} &  0 & \sqrt{3}  \\
0 & \sqrt{\frac{3}{2}} & 0 & -\frac{7}{2} & 0  \\
0 & 0 & \sqrt{3} & 0 & -\frac{9}{2}
\end{pmatrix}
\quad .
\end{equation}
The condition number of ${\bf D}^1$ is larger than one if its size is larger than $2 \times 2$, and the condition number increases with the size of the matrix. For a $56 \times 56$-sized matrix ${\bf D}^1$, as required for the computation of the first derivative of \gaia~DR3 spectra from 55 coefficients, the condition number is 66.2. As any derivative larger than one is a power of ${\bf D}^1$, the condition numbers of the derivative matrices increase with the power of the number of derivative. If the coefficients of a linear combination of Hermite functions are affected by noise, the relative noise on the coefficients for higher derivatives consequently increases strongly. It is therefore advisable to work with derivatives of as low order as possible in the case of noisy data. The highest derivatives we use in the course of this work is the fourth.\par
With the $k-$th derivative of a linear combination of Hermite functions being a linear combination of Hermite functions, the coefficients of which can be easily computed, we now proceed to the computation of the roots of a linear combination of Hermite functions.\par

\subsection{Roots of linear combinations of Hermite functions and their uncertainties}
From the relations in Eq.~(\ref{eq:rec1}) we see that the $n-$th Hermite function is actually a polynomial of degree $n$, multiplied by a Gaussian function ${\rm exp}\left\{-x^2/2\right\}$. Since the Gaussian function is a strictly positive function, it does not affect the position of the root of the Hermite function. The roots of the Hermite functions  coincide with the roots of the polynomial. Thus, a linear combination of Hermite functions with maximum order $N$ has exactly $N$ roots, which may be real and complex. For the computation of the actual roots, we may also follow the lines of root finding for polynomials. For polynomials in monomial representation, a standard approach to this problem is interpreting the polynomial as the characteristic polynomial of some matrix, called the companion matrix of the polynomial. With finding this matrix, the problem of root finding for a linear combination of monomials is transformed into an eigenvalue problem \citep{Press2007}, which can take advantage of the variety of fast and robust numerical eigenvalue techniques. For non-monomial representations of polynomials, an elegant generalisation of this approach has been developed for orthogonal polynomials. This approach makes use of the recurrence relationship for orthogonal polynomials, rather than the polynomial properties directly, and is therefore immediately transferable to Hermite functions, satisfying a suitable recurrence relationship expressed in Eq. (\ref{eq:rec1}). Here, we follow the line discussed by \cite{Day2005}, with the trivial transfer from orthogonal polynomials to Hermite functions. We only present the required results here and avoid the repetition of the not too complicated proofs and derivations, as it saves space and as they are presented by \cite{Day2005} in detail already.\par
We are considering a linear combination of Hermite functions according to Eq.~(\ref{eq:linearCombination}). With ${\bf \Phi}_N(x)$ we denote the $N-$ element vector of functions, containing the first $N$ Hermite functions, with indices from $0$ to $N-1$. Furthermore, with ${\bf c}_N$ we denote the vector of the first $N$ entries of the coefficient vector $\bf c$, and with ${\bf e}_N$ the $N$ element vector with a value of one as the $N-$th entry, and zero else. I.e., we have
\begin{eqnarray}
{\bf \Phi}_N(x) & \coloneqq & \left[\, \varphi_0(x), \varphi_1(x),\ldots,\varphi_{N-1}(x)\, \right]^{\mathsf T} \quad , \\
{\bf c}_N  & \coloneqq & \left[\, c_0, c_1,\ldots,c_{N-1}\, \right]^{\mathsf T} \quad , \\
{\bf e}_N   & \coloneqq & \left[\, 0, 0, \ldots, 0,1\, \right] ^{\mathsf T }\quad .
\end{eqnarray}
We assume $c_N \ne 0$ here, which can be ensured by trimming the length of $\bf c$ such that the highest order entry is non-zero if necessary. With these conventions, the function $\phi(x)$ from Eq.~(\ref{eq:linearCombination}) is written as
\begin{equation}
\phi(x) = {\bf \Phi}_N^{\mathsf T}(x) \, {\bf c}_N + c_N\, \varphi_N(x) \quad .
\end{equation}
For the moment it is not necessary to assume that the functions $\varphi_n(x)$ are actually the Hermite functions, we may only assume that these functions are satisfying a general recurrence relation of the form
\begin{equation}
x\, \varphi_{n-1}(x) = \sum \limits_{i=0}^n\, \varphi_i(x)\, h_{i,n-1} \quad . \label{eq:recGen}
\end{equation}
Then we can form the $N \times N$ matrix ${\bf H}$, whose elements are the $h_{i,j}$ from the recurrence relation above, with $i$ and $j$ running from $0$ to $N-1$. We define the $N \times N$ matrix $\bf B$, denoted the nonstandard companion matrix, by
\begin{equation}
{\bf B} \coloneqq {\bf H} -  \frac{h_{N,N-1}}{c_{N}} \, {\bf c}_N\, {\bf e}_N^{\mathsf T} \quad . \label{eq:nscm}
\end{equation}
A value $x_0$ is a root of the linear combination of the first $N+1$ functions $\varphi_n(x)$ with coefficient vector $\bf c$ if and only if $x_0$ is an eigenvalue of the matrix $\bf B$. Thus, the problem of root finding is again transformed into an eigenvalue problem for linear combinations of functions satisfying a suitable recurrence relation.\par
For Hermite functions, the general recurrence relation in Eq.~(\ref{eq:recGen}) takes the form given by Eq. (\ref{eq:rec1}), and we find
\begin{equation}
h_{i,j} =
\begin{cases}
	\sqrt{\frac{i+1}{2}} & \text{ for } j = i+1\\
	\sqrt{\frac{i}{2}} & \text{ for } j = i-1\\
	0 & \text{else} \quad .
\end{cases}
\end{equation}
As an example, for the case of $N=5$, the matrix $\bf B$ takes the form
\begin{equation}
{\bf B} =
\begin{pmatrix}
0 & \sqrt{\frac{1}{2}} & 0 & 0 & -\sqrt{\frac{5}{2}} \frac{c_0}{c_5} \\
\sqrt{\frac{1}{2}} & 0 & \sqrt{\frac{2}{2}} & 0 &  -\sqrt{\frac{5}{2}} \frac{c_1}{c_5} \\
0 & \sqrt{\frac{2}{2}} & 0 &  \sqrt{\frac{3}{2}} &  -\sqrt{\frac{5}{2}} \frac{c_2}{c_5}  \\
0 & 0 & \sqrt{\frac{3}{2}} & 0 &\sqrt{\frac{4}{2}} -\sqrt{\frac{5}{2}} \frac{c_3}{c_5} \\
0 & 0 & 0 & \sqrt{\frac{4}{2}} &  -\sqrt{\frac{5}{2}} \frac{c_4}{c_5}
\end{pmatrix} \quad .
\end{equation}
The nonstandard companion matrix $\bf B$ is thus easily generated for any given coefficient vector $\bf c$. The matrix $\bf B$ is of upper Hessenberg form, and the computations of the eigenvalues of a real Hessenberg matrix is particularly fast and simple (see, e.g., chapter 7.4 in \citealt{Golub2013}, or \citealt{NRwebnote17}).\par
We may however consider the case in which the coefficient vector $\bf c$ is not exact, but comes with an error, expressed in a covariance matrix $\Sigma_c$. For finding an estimate on the uncertainties of the roots of the linear combination of Hermite functions, we again may apply the approach developed for polynomial roots. We follow the formulations in Sect. 3.2.2 of \cite{Stetter2004} here, again with trivial transfer of argument from polynomials to Hermite functions.\par
The root of a linear combination $\phi(x)$ as in Eq.~(\ref{eq:linearCombination}) is a non-linear function of the coefficients in the vector $\bf c$. A linearisation of the problem is thus required for an efficient computation of the uncertainties of the roots. The roots of $\phi(x)$ we denote $z_i$, $i=1,\ldots,N$, and we consider the $z_i$ functions of the coefficient vector $\bf c$, as we do for the linear combination $\phi(x)$. We thus write
\begin{equation}
\phi \left( z_i({\bf c}),{\bf c}\right) = 0 \quad ,
\end{equation}
and compute the derivative of this expression with respect to $\bf c$, resulting in
\begin{equation}
\phi^{(1)}\left(z_i({\bf c}),{\bf c}\right) \cdot \frac{{\rm d}\,z({\bf c})}{{\rm d}{\bf c}} + \frac{\partial\,\phi(z_i({\bf c}),{\bf c})}{\partial {\bf c}} = 0 \quad .
\end{equation}
We thus obtain for each component $c_j$ of $\bf c$ the derivative
\begin{equation}
\frac{\partial z_i}{\partial c_j} = - \frac{\varphi_j(z_i)}{\phi^{(1)}(z_i)} \quad ,
\end{equation}
which are the elements $i,j$ of the Jacobian of the roots with respect to the coefficients of the linear combination, which we denote $\bf J$. In linear approximation, we obtain the covariance matrix of the roots, $\Sigma_z$, as
\begin{equation}
\Sigma_z = {\bf J} \, \Sigma_c \, {\bf J}^{\mathsf T} \quad .
\end{equation}

\subsection{Significance of local extrema \label{sec:significance}}

When considering a linear combination of Hermite functions, with coefficients affected by normally distributed random noise, we want to know if a given local extremum is statistically significant, or if it is a noise-induced feature that can be removed within some boundaries given by the covariance matrix on the coefficients. A local extremum is removed if the second derivative at the extremum is zero. In this case, the extremum is transformed to an inflection point. We might therefore consider the second derivate at the local extremum, normalised to its error, and the probability with which this value is zero. This probability might tell us about the statistical significance of the local extremum.\par
\begin{figure}
   \centering
   \includegraphics[width=0.49\textwidth]{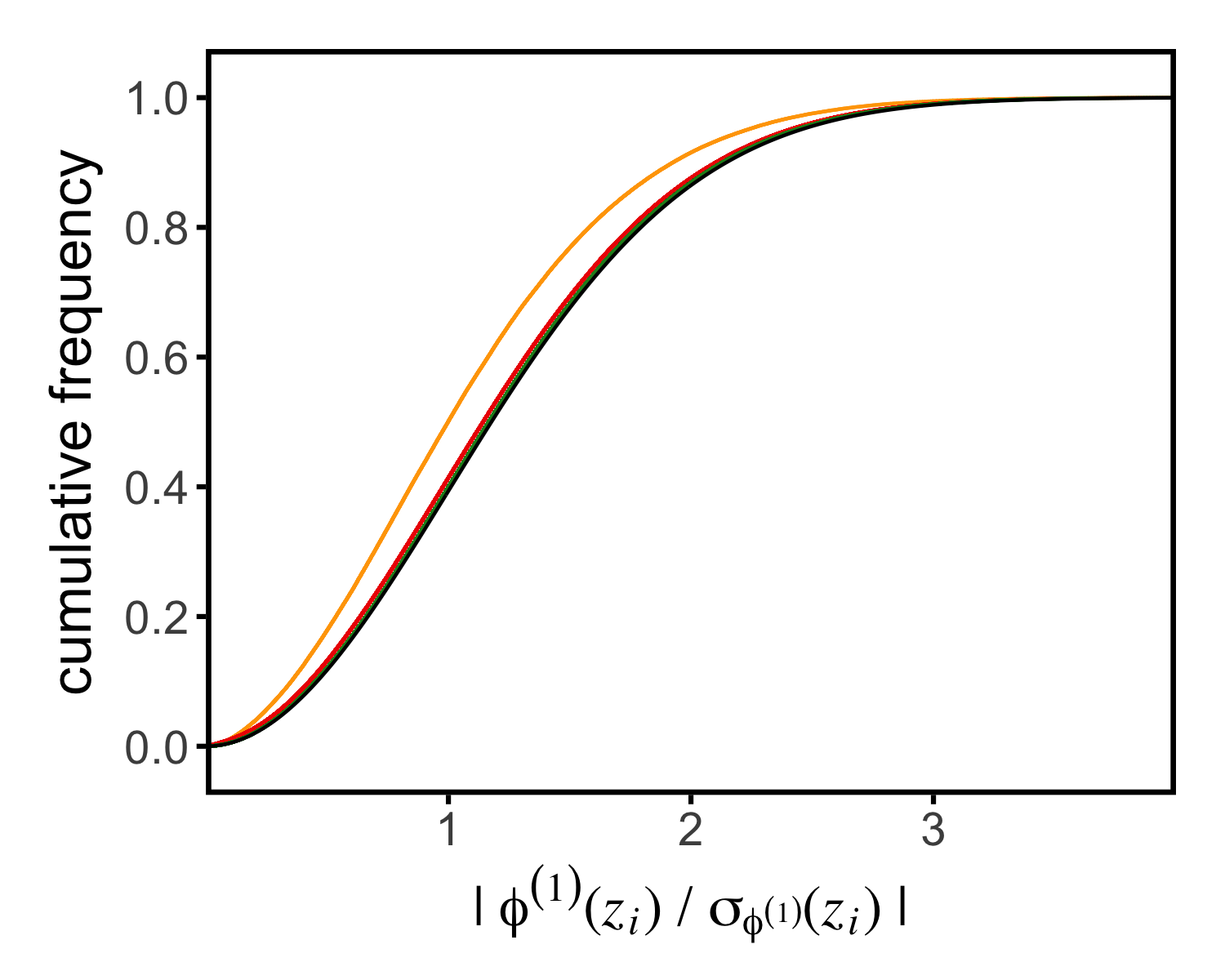}
   \caption{Empirical cumulative distribution functions for the absolute value of the derivative, normalised to its error, at the roots of linear combinations of Hermite functions, for $N=$ 5 (orange), 55 (red), and 200 (green). The black curve shows the cumulative distribution function for a chi distribution with two degrees of freedom.}
              \label{fig:noise0}
    \end{figure}

To investigate this approach quantitatively, we used Monte Carlo simulations. We generated random covariance matrices and for each of the random covariance matrices, we drew a large number of sets of random coefficients with zero mean. We then computed the roots of the corresponding random linear combination of Hermite functions, and the values of the derivative, divided by the error of the value of the derivative, at the roots. We then investigated the distribution function. We did this procedure for a variety of different numbers of coefficients. Figure~\ref{fig:noise0} shows resulting empirical cumulative distribution functions as examples, for different numbers $N$, and $5 \times 10^5$ random realisations. These results only include roots within the outermost extrema of the highest order Hermite function. With increasing $N$, the distribution approaches a chi distribution with two degrees of freedom.\par
With the cumulative distribution function of a chi distribution with two degrees of freedom, we can therefore estimate the significance of a local extremum of a linear combination of Hermite functions at $z_i$ by attributing an approximate $p$-value given by
\begin{equation}
 p(z_i) \approx 1 - \exp\left\{ - \frac{1}{2} \left( \frac{\phi^{(2)}(z_i)}{\sigma_{\phi^{(2)}(z_i)}} \right)^2 \right\} \quad .
\end{equation}

\subsection{Linear substitutions}
We close this section on the mathematical tools with a remark on the substitution $x \rightarrow (u-\Delta\theta)/\Theta$ in the argument of the Hermite functions. Such a transformation has been applied to \gaia~DR3 low resolution spectra to adjust the widths of the Hermite functions to the widths of the spectra \citep{DeAngeli2022}. In this case,
\begin{equation}
\int\limits_{-\infty}^{\infty} \varphi_n\left( \frac{u-\Delta\theta}{\Theta} \right) \varphi_m\left( \frac{u-\Delta\theta}{\Theta} \right) \, {\rm d}u = \Theta \, \delta_{nm} \quad.
\end{equation}
To maintain orthonormality, the Hermite functions after linear substitution need thus to by multiplied by $1/\sqrt{\Theta}$. The derivative of a Hermite function after substitution is
\begin{equation}
\frac{{\rm d}}{{\rm d}u}\, \varphi_n\left( \frac{u-\Delta\theta}{\Theta} \right) = \frac{1}{\Theta} \, \frac{{\rm d}}{{\rm d}x} \varphi_n(x) \quad .
\end{equation}
For the $k$-th derivative with respect to $u$, an additional factor $\Theta^{-k}$ thus needs to be applied to the result.\par
Taking the additional factors depending on $\Theta$ into account when differentiating and integrating Hermite functions with respect to $u$, we can use all results derived for linear combinations of Hermite functions also to Hermite functions after a linear substitution of its argument.\par
For general substitutions in the work of the form $f(x) \rightarrow f(y)$, we write $f(y)$ for $f(y(x)) \, {\rm d}y/{\rm d}x$.

\FloatBarrier

\section{Elements of the formation of spectra \label{sec:spectralFormation}}

Before entering into the analysis of spectral lines, we first resume some aspects of the formation of the spectra in this section, in order to establish the relationship between the low-resolution observational \gaia~spectrum and the underlying SPD. In particular we work out the role of the LSF, and the advantages resulting from their efficient parameterisation.\par
Often an observational spectrum $f(u)$ is related to the SPD $s(\lambda)$ by an integral transformation of the form
\begin{equation}
f(u) = \int \limits_{0}^{\infty} L(u,\lambda) \, R(\lambda) \, s(\lambda) \, \D{\lambda} \quad , \label{eq:spectrum}
\end{equation}
or a modified formulation thereof. Here, $L(u,\lambda)$ denotes the LSF, including the dispersion, and $R(\lambda)$ is the response function. In optical theory, the LSF is the inverse Fourier transform of the optical transfer function, integrated over the spatial direction \citep{Goodman1996}. For \gaia~DR3 low resolution spectra, this concept of an LSF is however not appropriate. The BP and RP spectra are obtained as a finite linear combination of Hermite functions constructed in the internal self-calibration process \citep{Carrasco2021}, and are not directly related to an optical transfer function describing the actual spectroscopic instrument. However, the XP spectra can still be linked to the SPD via an integral transformation of the general form of Eq.~(\ref{eq:spectrum}) if the function $L(u,\lambda)$ is given a more specific form. Setting
\begin{equation}
L(u,\lambda) \coloneqq \sum_{n=0}^{N} l_n(\lambda) \, \varphi_n(u) \quad ,
\end{equation}
putting this in Eq.~(\ref{eq:spectrum}) and using Eq.~(\ref{eq:coefficients}), we obtain for \gaia~DR3 low resolution spectra
\begin{equation}
f(u) \equiv \sum\limits_{n=0}^N c_n \varphi_n(u) = \int\limits_0^{\infty} L(u,\lambda) \, R(\lambda) \, s(\lambda) \, {\rm d}\lambda
\end{equation}
and
\begin{equation}
c_n = \int\limits_0^{\infty} l_n(\lambda)\, R(\lambda) \, s(\lambda) \, {\rm d}\lambda \quad . \label{eq:directCoefficients}
\end{equation}
Thus, the function $L(u,\lambda)$ is taking the role of the LSF, in the sense that it represents the instrument's response to a monochromatic signal. It is however not directly linked to the optical transfer function. The most important consequence of this is that $L(u,\lambda)$ is not necessarily a positive function. The response to a monochromatic input, when represented by a finite linear combination of Hermite functions, might show oscillational behaviour and be partly negative as well. We refer to $L(u,\lambda)$ as the ''pseudo-LSF'' in the following for convenience. The vector of all $l_n(\lambda)$, $n=0,\ldots,N$, we denote ${\bf l}(\lambda)$.\par
For the course of this work it will turn out convenient to allow for a fast computation of the vector ${\bf l}(\lambda)$, and we therefore consider developing $L(u,\lambda)$ not only in Hermite functions in the dependency on $u$, but also developing the wavelength dependency in basis functions. Doing so requires a suitable choice for the basis functions describing the wavelength dependency. For finding a good choice for the basis in wavelength, we note that we can write
\begin{equation}
l_n(\lambda) = \int\limits_{-\infty}^{\infty} L(u,\lambda) \, \varphi_n(u) \, {\rm d}u \quad.
\end{equation}
This is essentially applying the smoothing effect of the pseudo-LSF $L(u,\lambda)$ on the Hermite function $\varphi_n(u)$. If the dispersion relation linking $u$ and $\lambda$ is applied to substitute $\lambda$ with $u$, then the resulting function $l_n(u)$ might be very similar to the $n-$th Hermite function, in particular for small $n$. We therefore expect an efficient development of the function $L(\lambda,u)$ when substituting the wavelength with the pseudo-wavelength, which we denote $u^\prime$ to distinguish it from $u$, and use Hermite functions for the development in $u^\prime$ as well. We thus assume in good approximation
\begin{equation}
L(u,u^\prime) \approx \sum \limits_{n=0}^N\, \sum \limits_{m=0}^M \, \Lambda_{n,m}\, \varphi_n(u)\, \varphi_m(u^\prime) \quad , \label{eq:LSFapproximation}
\end{equation}
with $M$ only moderately larger than $N$. With $\boldsymbol \Lambda$ the matrix whose elements are the $\Lambda_{n,m}$, we thus obtain
\begin{equation}
{\bf l}(u^\prime) = {\boldsymbol \Lambda}\, {\bf \Phi}(u^\prime) \quad , \label{eq:LSFdevelopment3}
\end{equation}
with
\begin{equation}
{\bf \Phi}(u^\prime) \coloneqq \left[\, \varphi_0(u^\prime), \varphi_1(u^\prime),\ldots,\varphi_M(u^\prime) \, \right]^{\mathsf T} \quad .
\end{equation}
We make use of Eq.~(\ref{eq:LSFdevelopment3}) in Sect. \ref{sec:inversion} of this work. As a further preparation for Sect.~\ref{sec:inversion}, we consider the computation of some integrals involving the pseudo-LSF over $u^\prime$, namely integrals of the form
\begin{equation}
{\mathcal L}_{k,l}(u_0,u_1) \coloneqq \int \limits_{-\infty}^{\infty} L^{(k)}(u_1,u^\prime) \cdot (u^\prime - u_0)^{l} \, \D{u^\prime} \quad , \label{eq:LSFintegrals}
\end{equation}
for $k,l=0,1,2,\ldots$. Since the pseudo-LSF has an exponential decrease, these integrals exist. The choice of Hermite functions for the description of the dependency on both, $u$ and $u^\prime$, allows us to make use of results from Sect.~\ref{sec:mathematicalProperties} to find a simple solution to these integrals. For the case of $l=0$, from combining Eqs.~(\ref{eq:integral}), (\ref{eq:derivative}), and (\ref{eq:LSFdevelopment3}), we obtain immediately 
\begin{equation}
{\mathcal L}_{k,0}(u_0,u_1) = {\bf i}^{\mathsf T} \, {\bf D}^{\,k} \, {\boldsymbol \Lambda}^{\mathsf T} \, {\bf \Phi}(u_1) \quad . 
\end{equation}
For the case of $l=1$ we use the recurrence relation from Eq. (\ref{eq:rec1}) to develop the linear combination of Hermite functions multiplied by $(u^\prime - u_0)$ in a linear combination of Hermite functions. The coefficients of this linear combination are obtained by multiplication with a tridiagonal matrix ${\bf P}(u_0)$, with the elements
\begin{equation}
{\bf P}_{i,j}(u_0)  =
\begin{cases}
	\; - u_0 & \text{ for } i = j\\
	\; \sqrt{\frac{i}{2}} & \text{ for } i-j = -1  \\
	\; \sqrt{\frac{j}{2}} & \text{ for } i-j = 1  \\
	\; 0 & \text{else} \quad .
\end{cases}
\end{equation}
The case $l > 1$ we then obtain by multiplying ${\bf P}(u_0)$ $l$ times with itself, for which we write ${\bf P}^{\,l}(u_0)$. We thus get for the general case
\begin{equation}
{\mathcal L}_{k,l}(u_0,u_1) = {\bf i}^{\mathsf T} \, {\bf P}^{\,l}(u_0) \, {\bf D}^{\,k} \, {\boldsymbol \Lambda}^{\mathsf T} \, {\bf \Phi}(u_1) \quad ,
\end{equation}
an equation we use in Sect.~\ref{sec:inversion}, and which is a nice demonstration how the recurrence relations (\ref{eq:rec1}) and (\ref{eq:rec2}) for Hermite functions allow to compute integrals like in Eq. (\ref{eq:LSFintegrals}) efficiently using matrix algebra.\par
We close this section with the simple observation that the linearity of the integration implies that, if $s(\lambda)$ can be expressed as the sum of two SPDs, say $s(\lambda) = s_c(\lambda) + s_l(\lambda)$, the corresponding observational spectrum is the sum of the observational spectra of $s_c(\lambda)$ and $s_l(\lambda)$,
\begin{eqnarray}
f(u) & = & \int \limits_{0}^{\infty} L(u,\lambda) \, R(\lambda) \, \left[ s_c(\lambda) + s_l(\lambda) \right] \, \D{\lambda} \\
 & = & f_c(u) + f_l(u) \quad . \label{eq:lineSeparation}
\end{eqnarray}
If $s_c(\lambda)$ and $s_l(\lambda)$ separate the SPD onto a contribution from a continuum SPD and a SPD of a spectral line, respectively, we can thus separate also the observational spectra into a continuum contribution and a line contribution. For convenient notation in later sections, we introduce the abbreviation
\begin{equation}
t(\lambda) \coloneqq R(\lambda)\, s(\lambda) \quad . \label{eq:Rs}
\end{equation}

\section{Inversion of observational spectra \label{sec:inversion}}

After having derived the relevant mathematical properties of linear combinations of Hermite functions in Sect.~\ref{sec:mathematicalProperties}, and having combined the results of this section with the observational spectra in Sect.~\ref{sec:spectralFormation}, we now finally have to link all results to the external spectrum, i.e., we have to consider the inversion of internally calibrated \gaia~spectra to externally calibrated SPDs. This inversion is required to reduce the smearing effect of the pseudo-LSF when quantifying the line strengths, as discussed later. The inversion of observational spectra is intrinsically a complex problem. However, as far as this work is concerned, we may make some simplifying assumptions that will allow us to derive a simple and efficient formalism. First, for the study of a spectral line, we require a reliable inversion of the observational spectrum only in the vicinity of the spectral line. We do not need a reliable inversion of the entire observational spectrum. Far from the position of the line, the inversion might be arbitrarily wrong for our purposes. Second, we are mainly interested in the inversion of the continuum spectrum, where the spectral lines are removed. For the continuum we may well assume smoothness on the inverted spectrum. We are thus left with the problem of finding a local inversion to a smooth external spectrum, which is considerably easier than the problem of finding a global inversion to a general external spectrum.\par
A smooth local approximation to the inverted spectrum calls for a polynomial approximation in the vicinity of the line position. Naturally, the use of a Taylor expansion around the line position follows. This approach we line out in the following as the first of two approaches to the local smooth inversion.\par
We consider the product of the response curve and the continuum flux at some value $u_0$. To obtain $s_c(u_0)\, R(u_0) \eqqcolon t_c(u_0)$, we use Eq.~(\ref{eq:spectrum}), with Eq.~(\ref{eq:Rs}) and transformed from wavelength to pseudo-wavelength, which is
\begin{equation}
f(u_0) = \int\limits_{-\infty}^{\infty} L(u_0,u^\prime)\, t_c(u^\prime)\, \D{u^\prime} \quad .
\end{equation}
We introduce a Taylor expansion of $t(u^\prime)$ centred on $u_0$, resulting in
\begin{eqnarray}
f(u_0) & = & \int\limits_{-\infty}^{\infty} L(u_0,u^\prime) \, \left[ \sum\limits_{l=0}^{\infty} \frac{t_c^{(l)}(u_0)}{l!}\left( u^\prime - u_0 \right)^l \right] \, \D{u^\prime} \\
 & = & \sum\limits_{l=0}^{\infty} \frac{t_c^{(l)}(u_0)}{l!} \,  \int\limits_{-\infty}^{\infty} L(u_0,u^\prime) \, \left( u^\prime - u_0\right)^l\, \D{u^\prime} \\
 & = & \sum\limits_{l=0}^{\infty} \frac{t_c^{(l)}(u_0)}{l!} \, {\mathcal L}_{0,l}(u_0,u_0) \quad ,
\end{eqnarray}
with ${\mathcal L}_{k,l}(u_0,u_0)$ as defined in Eq.~(\ref{eq:LSFintegrals}). If we wish to approximate $t_c(u_0)$ in higher order approximation, we need to impose additional conditions. We can derive these conditions from higher derivatives at the same point $u_0$.\par
We compute the $k-$th derivative of $f(u)$ with respect to $u$, and obtain an analogous expression,
\begin{equation}
f^{(k)}(u_0) = \sum\limits_{l=0}^{\infty} \frac{t_c^{(l)}(u_0)}{l!} \, {\mathcal L}_{k,l}(u_0,u_0) \quad . \label{eq:firstApproach}
\end{equation}
If we truncate the Taylor expansion at the $K-$th term, we obtain a matrix equation
\begin{equation}
{\boldsymbol {\mathfrak f}}(u_0) = {\boldsymbol {\mathfrak L}}(u_0) \, {\boldsymbol {\mathfrak t}}(u_0) \quad , \label{eq:localInversion}
\end{equation}
with
\begin{eqnarray}
{\boldsymbol {\mathfrak f}}(u_0) &=& \left[\, f^{(0)}(u_0)\;f^{(1)}(u_0)\;f^{(2)}(u_0)\; \ldots \;f^{(K)}(u_0) \, \right]^{\mathsf T} \quad , \\ \label{eq:f}
{\boldsymbol {\mathfrak t}}(u_0) &=& \left[\, t_c^{(0)}(u_0) \;t_c^{(1)}(u_0)\;t_c^{(2)}(u_0)\;\ldots \; t_c^{(K)}(u_0) \, \right]^{\mathsf T} \quad ,
\end{eqnarray}
and the $(K+1) \times (K+1)$ matrix ${\boldsymbol {\mathfrak L}}(u_0)$, whose elements are
\begin{equation}
{\boldsymbol {\mathfrak L}}_{i,j}(u_0) = \frac{\mathcal L_{(i-1),(j-1)}(u_0,u_0)}{(j-1)!} \quad . \label{eq:L}
\end{equation}
Equation~(\ref{eq:localInversion}) is a formulation for the inversion of the observational spectrum by polynomial approximations to the product of the response function and the SPD.\par
The matrix $\boldsymbol {\mathfrak L}(u_0)$ has a simple form, namely
\begin{eqnarray}
{\boldsymbol {\mathfrak L}}_{i,j}(u_0)  & = & 0 \; \text{for } i > j \label{eq:prop1}  \\
{\boldsymbol {\mathfrak L}}_{i,i+1}(u_0) & =  &{\boldsymbol {\mathfrak L}}_{i+1,i+2}(u_0) \; \text{for } i \ge 0 \quad . \label{eq:prop2}
\end{eqnarray}
Equation (\ref{eq:prop1}) states that the matrix ${\boldsymbol {\mathfrak L}}$ is of upper triangular type. Equation~(\ref{eq:prop2}) states that the entries of the diagonal and the secondary diagonals are all identical, respectively. This implies that the $(K+1) \times (K+1)$ matrix ${\boldsymbol {\mathfrak L}}$ has only $K+1$ independent elements that need to be computed. This  allows for a fast computation of the matrix and its inverse, and makes the usage computationally simple.\par
For the computation of the formal error on $t_c(u_0)$, which we can use to judge the quality of the inversion of Eq.~(\ref{eq:localInversion}), it is convenient to write the vector $\boldsymbol {\mathfrak f}(u_0)$ as
\begin{equation}
{\boldsymbol {\mathfrak f}}(u_0) = {\bf H}(u_0) \, {\bf c} \quad ,
\end{equation}
with the $(K+1) \times (N+K)$ matrix ${\bf H}(u_0)$ whose $l$-th row is ${\bf \Phi}(u_0) \, {\bf D}^{(l-1)}$. Then,
\begin{equation}
{\boldsymbol {\mathfrak t}}(u_0) = {\boldsymbol {\mathfrak L}}^{-1} \, {\bf H}(u_0) \, {\bf c} \quad , \label{eq:localInversionFull}
\end{equation}
and the covariance matrix for ${\boldsymbol {\mathfrak t}}(u_0)$ is
\begin{equation}
\Sigma_{\boldsymbol {\mathfrak t}} = {\boldsymbol {\mathfrak L}}^{-1} \, {\bf H}(u_0) \, \Sigma_c \, {\bf H}^{\mathsf T}(u_0) \, \left( {\boldsymbol {\mathfrak L}}^{-1}\right)^{\mathsf T} \quad .
\end{equation}
The disadvantage of this method comes from the dependency on higher order derivatives of the spectrum at $u_0$. As discussed in the previous section, the signal-to-noise ratio of higher derivatives decreases strongly with increasing derivative, and local inversion becomes dominated by noise easily. In practice, only $K=1$ or $K=2$ might result in good estimates of $t_c(u_0)$.\par
It suggests itself here to replace the conditions on higher derivative by conditions on the zeroth derivative at different pseudo-wavelength positions, in order to become less sensitive to random noise in the inversion. The change is straight forward, replacing the derivative conditions in Eq.~(\ref{eq:firstApproach}) by conditions of $f^{(0)}(u_i)$ at different points $u_i$ in the vicinity of $u_0$. The results obtained with this approach however depend strongly on the choice of the $u_i$ and are not necessarily better than the results with derivative conditions. A more suitable approach results from adapting a technique for the global inversion from \cite{Weiler2020} to the problem of a local smooth inversion. We select a set of basis functions for the SPD, such as Legendre polynomials, and, assuming sufficient knowledge of the instrument, compute the observations spectra corresponding to the basis functions using Eq.~(\ref{eq:directCoefficients}). Locally approximating the observations spectrum around a pseudo-wavelength $u_0$ with these observational basis functions corresponds to approximating the SPD around a wavelength $\lambda(u_0)$ by Legendre polynomials. To avoid conditioning issues, we may adjust the argument of the Legendre polynomials to the range where these functions are orthonormal, i.e. the range $[-1,1]$. This approach is the same as used for the Hermite functions describing the \gaia~DR3 XP spectra. For sake of simplicity, we refer to this linear substitution for the Legendre polynomials $P_i(x)$ as $x = g(\lambda)$. For the approximation of the observational spectrum, we do not have to ensure an exact solution, but do a regression instead and reduce the influence of random noise. We choose an interval $\Delta$ in pseudo-wavelength, centred on $u_0$, and choose a grid of $M$ equidistant points on this interval. The grid point we arrange as a vector $\bf u$. Since the errors of the observational spectrum at the set of $M$ points $u_i$, $i=1,\ldots,M$ are likely to be correlated, we have to find a linear combination of observational spectra for the first $K$ Legendre polynomials, the coefficients of which are arranged as a vector $\bf b$, that minimise the squared Mahalanobis distance
\begin{equation}
D_m^2 = \left( f({\bf u}) - {\bf \Phi}^{\mathsf T}(u) {\bf L}\, {\bf b} \right)^{\mathsf T} \, \Sigma_f^{-1} \, \left( f({\bf u}) - {\bf \Phi}^{\mathsf T}({\bf u}) {\bf L}\, {\bf b} \right) \quad .
\end{equation}
Here, $\bf L$ is the $N \times K$ matrix of coefficients of the observational spectra corresponding to the first $K$ Legendre polynomials, developed in Hermite functions. This is a generalised least squares problem with the solution \citep{Kariya2004}
\begin{equation}
{\bf b} = \left( \left({\bf \Phi}^{\mathsf T}({\bf u}) {\bf L}\right)^{\mathsf T} \Sigma_f^{-1} \left({\bf \Phi}^{\mathsf T}({\bf u}) {\bf L}\right) \right)^{-1} \, \left({\bf \Phi}^{\mathsf T}({\bf u}) {\bf L}\right)^{\mathsf T} \Sigma_f^{-1} \, f({\bf u}) \quad . \label{eq:generalisedLeastSquares}
\end{equation}
The linear combination
\begin{equation}
s(\lambda_0) = \sum\limits_{i=0}^{K-1}\, b_i \cdot P_i(g(\lambda_0)) \equiv {\bf P}_L(\lambda_0) \, {\bf b}  \label{eq:legendreInversion}
\end{equation}
is then an approximation to the SPD at wavelength $\lambda_0$. The results of the inversion of the observational spectrum with this approach depend in the choice of $\Delta$, $K$, and $M$. A good guideline for the choices is $M > K$, and the conditioning of the matrix to be inverted in Eq.~(\ref{eq:generalisedLeastSquares}) being small enough to be well off any numerical problems. For the tests done in this work we used $K=5$, $\Delta$ around 6, and $M$ around 15.\par
Both methods for local smooth inversion of observational spectra have their advantages and disadvantages. The great advantage of the second approach, the inversion using Legendre polynomials, is the lower sensitivity to random noise. This approach however requires the explicit use of the response functions, and systematic errors in the response function immediately affect the inversion. This is not the case of the first approach, which is solving for the product of the external spectrum and the response function directly. As we will see in Sec.~\ref{sec:narrowLines}, this is an advantage in the treatment of narrow spectral lines. So in general the second approach might be more advisable. In cases of strong lines and good signal to noise on the observational spectrum, potential systematic errors on the response function might become larger than the random noise, and the first approach might be more suitable. We will re-use the second approach in Sect.~\ref{sec:continuumBroad} not for inversion, but for continuum estimation.

\section{Spectral lines as local extrema \label{sec:localExtrema}}

A spectral line might result in a local extremum of the observational spectrum $f(u)$. In particular we assume that an emission line results in a local maximum, and an absorption line in a local minimum. A necessary condition for a local extremum of $f(u)$ is a root of $f^{(1)}(u)$. The position of such a root of the first derivative we denote with $u_E$. For $f^{(2)}(u_E) < 0$ the observational spectrum has a local maximum at $u_E$, and for $f^{(2)}(u_E) > 0$ a local minimum.  We omit higher order derivative tests for simplicity here.\par
A local extremum is however not necessarily caused by a spectral line, as also noise in the coefficients representing the spectra can manifest in wavy structures with many local extrema. As a local extremum requires a root in the derivative, noise is more likely to result in local extrema in regions of an observational spectrum where the first derivative of the true, noise-free spectrum is small, i.e. where the gradient of the spectrum is low. With decreasing signal-to-noise ratio, more and more regions with increasing gradient of the spectrum become likely to show local extrema due to noise. We therefore need an estimator quantifying the statistical significance of a local extremum, to distinguish between noise patterns and local extrema that actually are produced by the presence of a spectral line. For this purpose, we can use the value of the second derivative, normalised by their error, as discussed in Sect.~\ref{sec:significance}. For noise-induced local extrema, this normalised value of the second derivative at the position of an extremum follows approximately a chi-distribution with two degrees of freedom, from which we can compute a $p$-value for a local extremum.\par
\begin{figure}
   \centering
   \includegraphics[width=0.49\textwidth]{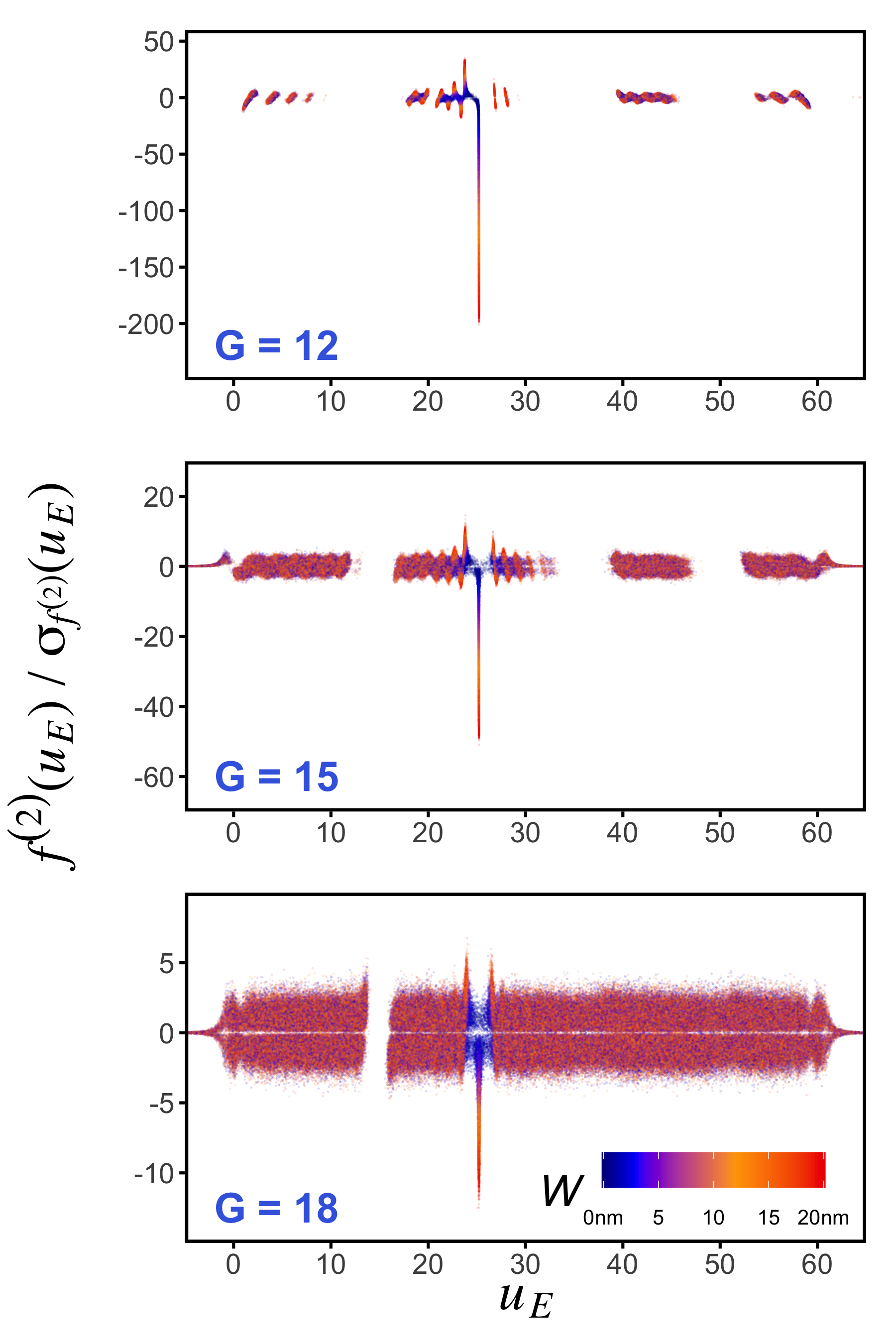}
   \caption{The error-normalised value of the second derivative of the BP spectrum as a function of $u_E$ for random spectra consisting of a $2 \times 10^4$~K Black Body continuum and a narrow H$\beta$ line with random equivalent width $W$. The three panels correspond to $G$-magnitudes of 12\m, 15\m, and 18\m, from top to bottom.}
              \label{fig:noise2}
    \end{figure}

The noise-induced local extrema are illustrated for simulated BP spectra in Fig.~\ref{fig:noise2}. Here, the value of the second derivative at the root of the first derivative, normalised to its error, is shown for simulated spectra. The simulations in this work are done with the simple \gaia~XP simulator described in \cite{Weiler2020}. The simulated spectra are composed of a Black Body continuum with $2 \times 10^4$~K, and a narrow H$\beta$ line with a random chosen equivalent width, $W$, which we define as
\begin{equation}
W \coloneqq \int\limits_{0}^{\infty} \frac{s(\lambda) - s_c(\lambda)}{s_c(\lambda)}\, \D{\lambda} \quad . \label{eq:defW}
\end{equation}
This definition of $W$ has the opposite sign than in usually used definitions (e.g. \citealt{Carroll2017}). We use this definition here to apply it to absorption and emission lines alike, and we prefer emission lines to have a positive equivalent width, and absorption lines to have a negative one. The result for $2 \times 10^5$ simulated BP spectra for $G$-band magnitudes for the continuum of 12, 15, and 18, respectively, are shown. The colour scale shows the random value of $W$ in each of the $10^5$ spectra.\par
With increasing magnitude, and thus decreasing signal-to-noise-ratio, the number of local extrema increases. For the lowest magnitude, these local extrema caused by noise form predominantly in parts of the spectrum where the gradient is small. With increasing magnitude, the noise-related local extrema spread out. For $G =18^{\rm m}$, the only part in the BP spectrum free of noise-related local extrema is a narrow region around the wavelength cut-off position, where the gradient of the spectrum is largest. The simulated H$\beta$ line shows very clearly against the noise background of local extrema. And the larger the equivalent width of the simulated H$\beta$ line, the smaller the value of the second derivative at the root of the first derivative, normalised to its error.\par
If the spectral line is weak, i.e. it has a small absolute value of $W$, and the slope of the spectrum is sufficiently large, an emission or absorption line might not result in a local extremum of the spectrum, though. A line might manifest itself in this case as a deformation of the spectrum. To detect such lines which are weak compared to the slope of the spectrum, we might consider higher order derivatives. The deformation of the spectrum results in a local extremum in curvature of the spectrum, i.e. in $f^{(2)}(u)$. In this work, we first discuss the case of a spectral line as a local extremum in $f(u)$ in detail. The following sections analyse this case for narrow and broad lines. Having worked out the necessary formalism for this case, it is easy to transfer the results to extrema in the curvature of the spectrum by simply increasing all derivatives involved by two. This latter case is analysed in detail in Sect.~\ref{sec:higherOrder}.

\section{The limit of narrow lines \label{sec:narrowLines}}

In this section we consider the case of narrow spectral lines, where we understand 'narrow' in the sense that the intrinsic width of the spectral line in the SPD (measured, e.g. by its FWHM) is small compared to the width of the LSF, expressed in wavelength. In this case we can neglect the width of the spectral line in the SPD, and write it as the sum of a continuum SPD and a weighted Dirac delta distribution,
\begin{eqnarray}
s(\lambda) & = &  s_c(\lambda) + s_l(\lambda)\\
 & \approx & s_c(\lambda) + \alpha \, \delta(\lambda-\lambda_L) \quad . \label{eq:narrowLine}
\end{eqnarray}
Here $s_c(\lambda)$ is the continuum SPD, $s_l(\lambda)$ the line SPD, and $\lambda_L$ is the wavelength of the line. Using Eq.~(\ref{eq:spectrum}) and using the substitution of $u^\prime$ for the wavelength $\lambda$, we obtain in this case for the observed spectrum
\begin{eqnarray}
f(u) & = & \int\limits_{-\infty}^{\infty} L(u,u^\prime)\, R(u^\prime) \, \left[\, s_c(u^\prime) + \alpha\, \delta(u^\prime-u_L)\, \right] \, \D{u^\prime} \\
 &= & f_c(u) + \alpha\, R(u_L) \, L(u,u_L) \label{eq:lineSeparation} \\
 & = & f_c(u) + f_l(u) \quad .
\end{eqnarray}
The observed spectrum in the limit of a narrow spectral line is thus the sum of the observed continuum spectrum $f_c(u)$ and a weighted LSF as the observed line $f_l(u)$.

\subsection{Systematic errors in line positions \label{sec:systematicErrorInPosition}}

As we are considering spectral lines as local extrema in the internally calibrated spectrum, we have to analyse the relationship between the position of the spectral line, $u_L$, and the position at which the corresponding local extremum is occurring, $u_E$. Using Eq.~(\ref{eq:lineSeparation}), we obtain for a local extremum the necessary condition
\begin{equation}
f^{(1)}(u_E) + \alpha\, R(u_L) \, L^{(1)}\left(u_E,u_L\right) = 0 \quad .
\end{equation}
We assume $\alpha \ne 0$, and for convenience we introduce the abbreviation
\begin{equation}
\zeta(u) \coloneqq - \frac{f^{(1)}_c(u)}{R(u_L)\, s_c(u_L)} \quad . \label{eq:zeta}
\end{equation}
At this place it is convenient to express the strength of the line with respect to the underlying continuum, and use the equivalent width as defined in Eq.~(\ref{eq:defW}). In the approximation of a narrow line, and using Eq.~(\ref{eq:narrowLine}), the equivalent width becomes
\begin{equation}
W = \frac{\alpha}{s_c(\lambda_L)} \quad . \label{eq:W2}
\end{equation}
Substituting with $u_L$, we obtain
\begin{equation}
L^{(1)}(u_E,u_L) = \frac{\zeta(u_E)}{W} \quad . \label{eq:linePosition}
\end{equation}
This is a strongly non-linear equation for the position of the local extremum, $u_E$, given the line position $u_L$. Although this equation is not useful for practical computations in relating $u_E$ and $u_L$, it nevertheless allows for a graphical illustration. In the limit of an extremely strong line, i.e. for the limit $W \rightarrow \infty$, the right hand side of Eq.~(\ref{eq:linePosition}) becomes zero, and the position of the extremum coincides with the position of the line. In the limit of negligible line strength, i.e. $W \rightarrow 0$, the right hand side becomes arbitrarily large, and no more solution to Eq.~(\ref{eq:linePosition}) may exist. In this case, the line is not introducing a local extremum in the observational spectrum anymore. Between these two extreme cases, the position of the local extremum is shifted with respect to the line position.\par
\begin{figure}
   \centering
   \includegraphics[width=0.49\textwidth]{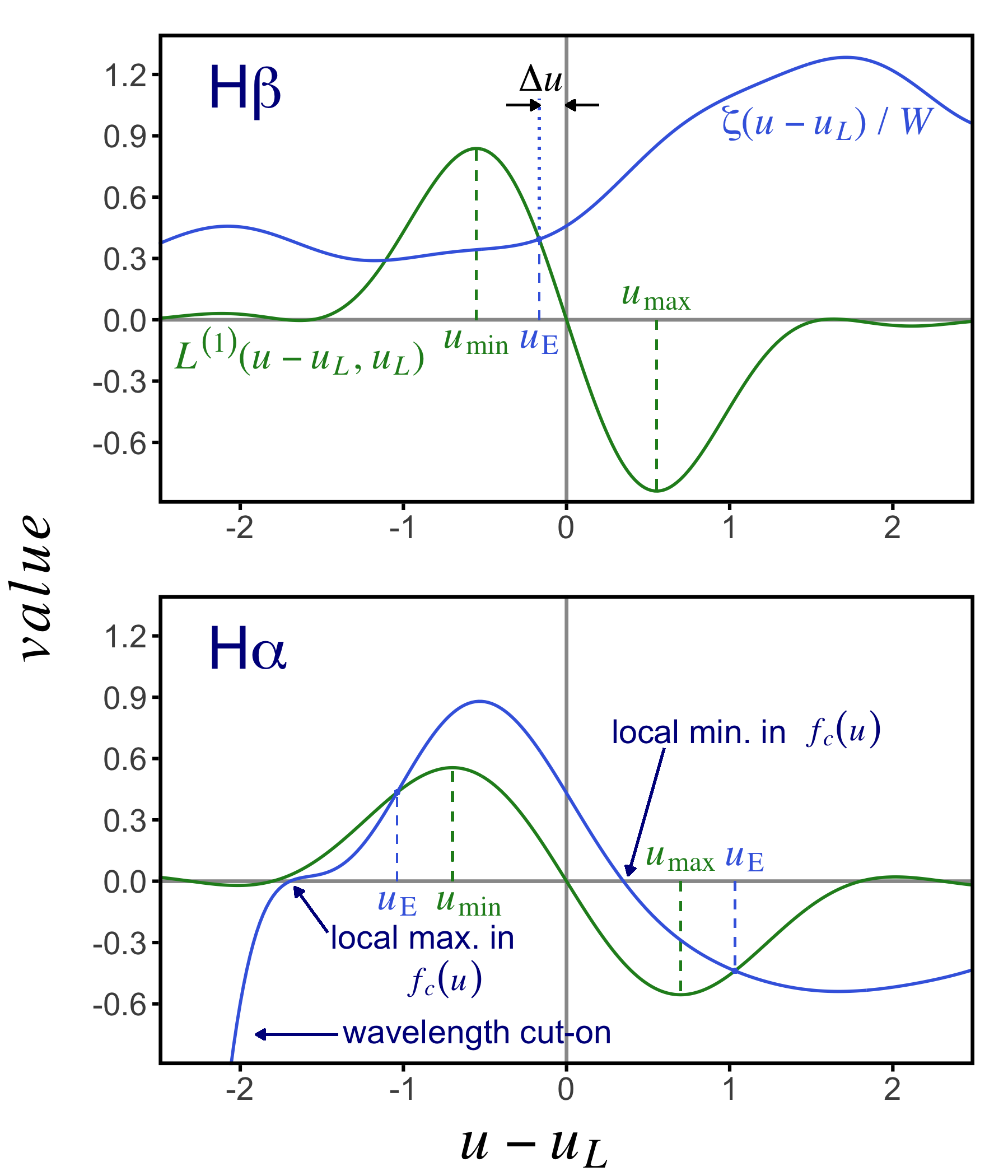}
   \caption{Illustration of the shift between the line position $u_L$ and the position of local extrema $u_E$ for a $2\times10^4$K Black Body background and an H$\beta$ line (top panel) and an H$\alpha$ line (bottom panel). The green lines show $L^{(1)}\left(u-u_L,u_L\right)$, the blue lines show $\zeta\left(u-u_L\right)$ divided by $W$. For more explanation see text.}
              \label{fig:systematicShift}
    \end{figure}

This is illustrated with a simulated spectrum, again consisting of a Black Body continuum of $2\times 10^4$K and narrow H$\alpha$ (in RP) and H$\beta$ (in BP) lines. The functions $L^{(1)}(u,u_L)$ and $\zeta(u)/W$ are shown in Fig.~\ref{fig:systematicShift}. A local extremum forms where the two functions intersect, this position is indicated $u_E$ in the figure. A good range to look for extrema associated with a line is the range between the first local extrema of $L^{(1)}(u,u_L)$. The shift $\Delta u$ is towards higher values in the underlying continuum spectrum for emission lines, and towards lower values for absorption lines. As a criterion for associating a local extremum with a line at known wavelengths, and thus with known $u_L$, we might therefore require a shift $\Delta u$ which is within the extrema of the first derivative of $L(u,u_L)$, and towards lower values for a local minimum, and towards larger values for a local maximum.\par
For H$\alpha$, the situation is more complicated. For a faint line, no local extremum within the expected range of $[u_{\rm min},u_{\rm max}]$ forms. Instead, two local extrema, i.e. a wavy-like structure, forms around the line position. This is a consequence of the more complex structure of the response curve close to the wavelength cut-on position of RP. This filter cut-on is realised by an interference filter on the RP prism, and interference filters tend to have a wavy transmission pattern in particular close to strong gradients as a cut-on flank. This pattern, which is present in the nominal response curves for the \gaia~instruments \citep{Jordi2010}, interfere with the H$\alpha$ line. Even if the nominal pre-launch response curves might not be fully representative for the calibrated instrument, such patterns, as they are based on physical properties of the instrument, might still occur and complicate the analysis of faint H$\alpha$ lines.\par
In practice, the function $\zeta(u_E)$ is unknown, and with it the shift $\Delta u$ between the position of the extremum and of the line position. However, we may expect the shift $\Delta u$ to be with the local extrema of the first derivative of the pseudo-LSF, at $u_{\rm min}$ and $u_{\rm max}$. With an assumption for the pseudo-LSF, we can compute these positions and consider a local extremum in agreement with a line position if $u_{\rm min} < u_E < u_{\rm max}$.

\subsection{Continuum model for narrow lines \label{sec:continuumNarrow}}

Having $f(u)$ available, we have to separate the continuum spectrum from the line contribution. This requires the construction of a continuum spectrum which is to be subtracted from the observational spectrum $f(u)$. To simplify this procedure, we use a simple approximation. We assume that the continuum spectrum is characterised by
\begin{equation}
f^{(2)}_c(u_E) = 0 \quad , \label{eq:continuumCondition}
\end{equation}
that is, we assume that the curvature of the observed continuum spectrum is zero at the maximum intensity of the spectral line. Differentiating Eq.~(\ref{eq:lineSeparation}) twice with respect to $u$ results in
\begin{equation}
f^{(2)}(u) = f^{(2)}_c(u) + \alpha\, R(u_L) \, L^{(2)}(u,u_L) \quad . \label{eq:continuumCondition2}
\end{equation}
With Eq.~(\ref{eq:continuumCondition}), this becomes
\begin{equation}
\alpha \, R(u_L) \approx \frac{f^{(2)}(u_E)}{L^{(2)}(u_E,u_L)} \quad . \label{eq:alphaR}
\end{equation}
and
\begin{equation}
f_c(u) \approx f(u) -  \frac{f^{(2)}(u_E)}{L^{(2)}(u_E,u_L)} \, L(u,u_L) \quad . \label{eq:continuum}
\end{equation}
With Eq.~(\ref{eq:continuumCondition}) as a condition for estimating the continuum spectrum, we can thus estimate the line spectrum from the ratio of the curvature of the observed spectrum at the position where the line is strongest in the observational spectrum, and the curvature of the LSF at the same position. In case the true line position, $u_L$ is not known, we may use the approximation $u_L \approx u_E$.\par
Equation~(\ref{eq:continuum}) translates with Eq.~(\ref{eq:LSFdevelopment3}) into an equation for the vector of continuum coefficients, ${\bf c}_c$,
\begin{equation}
{\bf c}_c \approx {\bf c} - \frac{f^{(2)}(u_E)}{L^{(2)}(u_E,u_L)} \, {\bf l}(u_L) \quad .
\end{equation}

\subsection{Equivalent widths for narrow lines \label{sec:WforNarrowLines}}
For a spectral line we are not interested in the strength of the line scaled with the response function, $\alpha\, R(u_0)$, as provided by Eq.~(\ref{eq:alphaR}), but rather in the equivalent width $W$. To obtain an expression for $W$, we use Eq.~(\ref{eq:W2}), with a substitution by $u_L$, and combine this equation with Eq.~(\ref{eq:alphaR}), to obtain
\begin{equation}
W = \frac{f^{(2)}(u_E) }{L^{(2)}(u_E,u_L)} \frac{1}{s_c(u_L)\, R(u_L)} \equiv \frac{f^{(2)}(u_E) }{L^{(2)}(u_E,u_L)} \frac{1}{t_c(u_L)}\quad . \label{eq:W1}
\end{equation}
For the determination of the equivalent width from the observational spectrum we thus need to derive $s_c(u_0)\, R(u_0)$ from the continuum spectrum provided by Eq.~(\ref{eq:continuum}). Here, we can apply the inversion approaches as discussed in Sect.~\ref{sec:inversion}. Equation (\ref{eq:localInversion}), or Eq.~(\ref{eq:legendreInversion}) together with a response function, can be combined with Eq.~(\ref{eq:W1}) to obtain the equivalent width of a spectral line.\par
\begin{figure}
   \centering
   \includegraphics[width=0.48\textwidth]{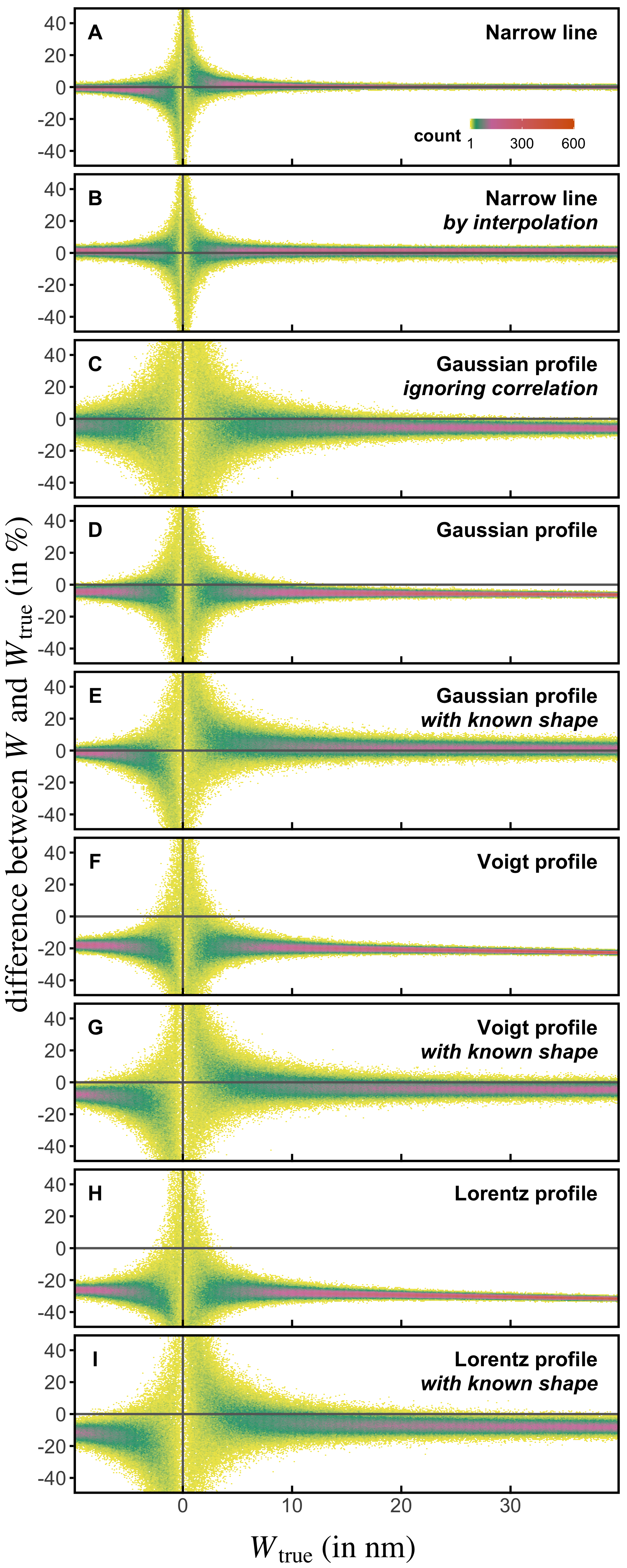}
   \caption{Distributions of the relative difference between the computed equivalent width ($W$) and the true equivalent width ($W_{true}$) of H$\beta$ lines with a $2 \times 10^4$ K Black Body continuum for different assumptions on the line shape (narrow lines, Gaussian, Voigt and Lorentz profiles) and different computation techniques. Each panel shows the result from $5 \times 10^5$ random realisations of a spectrum for $G=12^{\rm m}$ and 10 transits. Positive values in the horizontal axis correspond to H$\beta$ emission, negative values to H$\beta$ absorption. For details see text in Sects. \ref{sec:WforNarrowLines}, \ref{sec:WforBroadLines}, and \ref{sec:broadKnown}.}
              \label{fig:lineInterpolationExample}
    \end{figure}

We can compute the uncertainty on the equivalent width in linear approximation using the Jacobian of $W$ with respect to the coefficients of the XP spectrum. The elements of this are
\begin{equation}
\frac{\partial W}{\partial c_i} = \frac{1}{L^{(2)}(u_E,u_L)} \left[ \frac{1}{t_c(u_E)} \frac{\partial}{\partial c_i}f^{(2)}(u_E) - \frac{f^{(2)}(u_E)}{t_c^2(u_L)} \frac{\partial}{\partial c_i} t_c(u_L) \right] \quad .
\end{equation}
The derivative of $f^{(2)}(u_E)$ with respect to $c_i$ is the $i$-th element of the vector ${\bf \Phi}^{\mathsf T}(u_E) {\bf D}^{2}$. The expression for the derivative of $t_c(u_L)$ with respect to $c_i$ depends on the approach chosen for inversion. Using Eq.~(\ref{eq:localInversionFull}), the derivative is the $i$-th element of the first row of the matrix ${\boldsymbol {\mathfrak L}}^{-1} {\bf H} - {\boldsymbol {\mathfrak L}}^{-1} {\bf H}{\bf \Phi}^{\mathsf T}(u_E) {\bf D}^{2}\, {\bf l}(u_L)$. Using Eq.~(\ref{eq:generalisedLeastSquares}) and (\ref{eq:legendreInversion}), the derivative of $t_c(u_L)$ with respect to $c_i$ is the $i$-th entry of
\begin{equation}
\begin{split}
R(u_L) \, {\bf P}_L({\bf u}) \left( \left({\bf \Phi}^{\mathsf T}({\bf u}) {\bf L}\right)^{\mathsf T} \Sigma_f^{-1} \left({\bf \Phi}^{\mathsf T}({\bf u}) {\bf L}\right) \right)^{-1}  \left({\bf \Phi}^{\mathsf T}({\bf u}) {\bf L}\right)^{\mathsf T} \Sigma_f^{-1} \, {\bf \Phi}({\bf u}) \\  \left[1 - {\bf D}^{2}\,{\bf l}(u_L) \right]\; .
\end{split}
\end{equation}
A simple example computation for the illustration of the approach for narrow lines is shown in panel A of Fig.~\ref{fig:lineInterpolationExample}. As before simulations for a Black Body continuum for a temperature of $2\times 10^4$~K and a $G$-magnitude of 12$^{\rm m}$ was used, and a narrow H$\beta$ line was added, with a randomly chosen equivalent width. Then, the BP spectrum was computed and the coefficients of the Hermite representation computed from 10 transits. The equivalent width was computed from these coefficients, and from it the relative difference to the true equivalent width. This was repeated for $5 \times 10^5$ random realisations. The distribution of the relative difference for these simulations is shown as a function of the true equivalent width. In this computation, no a-priori information on the line position was used, i.e., the approximation $u_L = u_E$ was used. Even with this approximation, a very good determination of $W$ is possible, with systematic errors on the level of a per cent.

\subsection{Blended narrow lines \label{sec:blended}}

In the following, we are considering the case that two or more narrow lines are not sufficiently separated in pseudo-wavelength to neglect the overlap between them. We refer to this case as blending of lines. The approach applied here is very similar to the case of unblended narrow lines, we can still use the condition  expressed by Eq.~(\ref{eq:continuumCondition}) in this case. But we have to modify Eq.~(\ref{eq:continuumCondition2}) in such a way that it includes the contributions from other lines to a particular line under consideration. Taking $n_l$ lines into account, with the line fluxes $\alpha_i$, $i=1,\ldots,n_l$ and a pseudo-wavelength position $u_{L,i}$ for the $i$-th line, we obtain
\begin{equation}
f^{(2)}(u) = \sum\limits_{i=1}^{n_l}\alpha_i \, R\left(u_{L,i} \right) \,L^{(2)}\left(u,u_{L,i}\right) \quad .
\end{equation}
Evaluating this expression at the line positions $u_{L,j}$ and writing it as a matrix equation, we get
\begin{equation}
{\bf f}^{(2)}({\bf u}) = {\bf L}^{(2)}\, {\bf r}\left({\bf u}\right) \quad , \label{eq:blended}
\end{equation}
with ${\bf L}^{(2)}$ the matrix with the elements ${\bf L}^{(2)}_{i,j} = L\left(u_{L,i},u_{L,j} \right)$, ${\bf f}^{(2)}({\bf u}) $ the vector containing the elements $f^{(2)}\left(u_{L,i}\right)$ and ${\bf r}\left({\bf u}\right)$ the vector containing the elements $\alpha_i\, R\left(u_{L,i}\right)$, $i=1,\ldots,n_l$ in both cases. Equation~(\ref{eq:blended}) is then solved for ${\bf r}({\bf u})$, and the computation of the equivalent widths for the $n_l$ lines continues in the same way as discussed for the unblended case before.

\subsection{Upper limits for narrow lines \label{sec:upperLimits}}
In case of non-detection of a particular spectral line, it might be of interest to derive an upper limit for the equivalent width of the non-detected line from the coefficients of the \gaia~DR3 spectrum. Given the covariance matrix for the coefficients of the spectrum, $\Sigma_c$, and a statistical level of significance $p$, we can compute what line, given by $\alpha \, R(u_L)$, can be added to the spectrum within the level of significance. With the coefficients for the pseudo-LSF at the position of the line, ${\bf l}(u_L)$ given by Eq.~(\ref{eq:LSFdevelopment3}), the distance between a spectrum and a spectrum with a narrow line added, expressed in numbers of standard deviations, is given by the Mahalanobis distance
\begin{equation}
D_M = \left| \alpha R(u_L) \right| \, \sqrt{{\bf l}^{\mathsf T}(u_L) \, \Sigma_c^{-1}\, {\bf l}(u_L)} \quad.
\end{equation}
This statistics follows a chi-distribution with $n$ degrees of freedom. With $Q(p,n)$ the quantile function of the chi-distribution with $n$ degrees of freedom and the coefficients for the pseudo-LSF at the position of the line, ${\bf l}(u_L)$ given by Eq.~(\ref{eq:LSFdevelopment3}), we obtain
\begin{equation}
\left| \alpha_{limit} R(u_L) \right| = \frac{Q(p,n)}{\sqrt{{\bf l}^{\mathsf T}(u_L) \, \Sigma_c^{-1}\, {\bf l}(u_L)}} \quad .
\end{equation}
This result can be used with Eq.~(\ref{eq:W1}) to derive an upper limit on the absolute value of the equivalent width of a non-detected line,
\begin{equation}
W_{limit} = \frac{\left| \alpha_{limit} R(u_L) \right| }{t(u_L)} \quad .
\end{equation}
With the cumulative distribution function for the chi distribution, $P(n/2,p^2/2)$ with $P(k,x)$ the incomplete gamma function \citep{Press2007}, the 0.6827 quantile is approximately 7.7094, and the 0.95 and the 0.99 quantiles are approximately 8.5622 and 9.0715, respectively.

\section{Broad lines \label{sec:broadLines}}

\subsection{Detection of broad lines}

\begin{figure}
   \centering
   \includegraphics[width=0.49\textwidth]{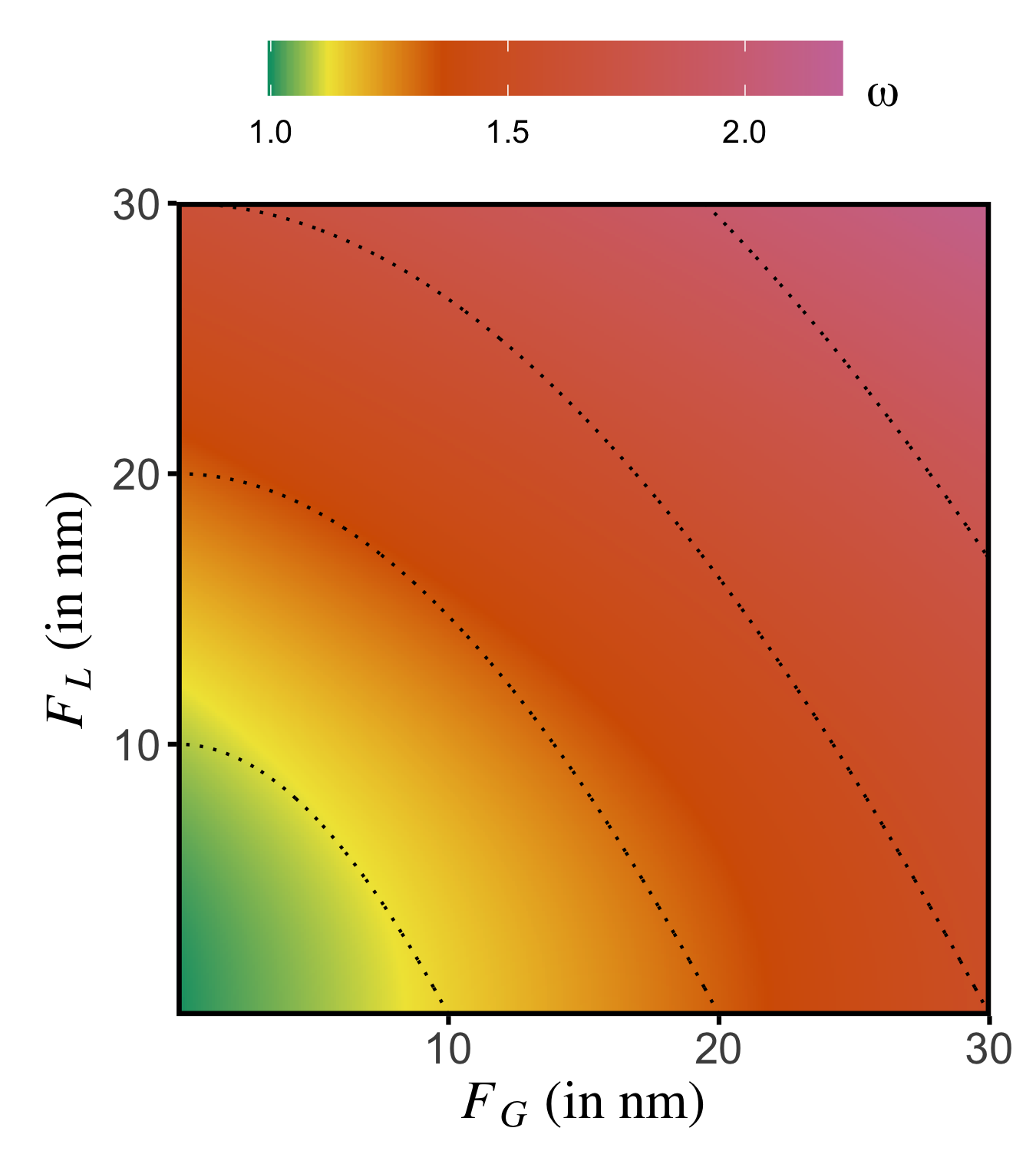}
   \caption{The line width parameter $\omega$ for Voigt profiles, as a function of the FWHM of the Gaussian contribution (on the horizontal axis) and the Lorentz contribution (on the vertical axis). The dotted lines indicate iso-contours in FWHM of 10, 20, 30, and 40 nm. For details see text.}
              \label{fig:lineShape}
    \end{figure}

We now consider broad lines, with which we mean that the approximation of the line by a Delta function is no longer suitable. Instead, the intrinsic width of the line is so large that it cannot be neglected even for the low spectral resolution of the \gaia~spectrophotometers. For the distinction between narrow and broad lines we require a measure for the line width. We can base this measure on roots of derivatives of the spectrum, if we consider inflection points that are related to a local extremum. A local extremum has an inflection point on either side, which we denote $u_{i,1}$ and $u_{i,2}$, and base the estimator for the line width on the distance between these two points. As inflection points are roots of the second derivative of the observational spectrum, for a narrow line the distance between the inflection points is in first order approximation the distance between the innermost inflection points of the pseudo-LSF. With increasing intrinsic line width, the distance between the inflection points increases with respect to the corresponding distance for the pseudo-LSF. We therefore normalise the distance between the innermost two inflection points of the pseudo-LSF, and use this dimensionless quantity, which we denote $\omega$,
\begin{equation}
\omega = \frac{ u_{i,2} - u_{i,1} }{ u_{i,2}^{[LSF]} - u_{i,1}^{[LSF]} } \quad . \label{omega}
\end{equation}
as the measure for line widths. If this quantity is significantly larger than one, we may consider the line broad. For the uncertainty on $\omega$, we have to take the correlations between the errors on $u_{i,1}$ and $u_{i,2}$ into account, which we obtain from the corresponding elements of the covariance matrix on the roots of the second derivative,
\begin{equation}
\sigma_\omega = \frac{\sqrt{\Sigma_{11} + \Sigma_{22} - 2\Sigma_{12}}}{ u_{i,2}^{[LSF]} - u_{i,1}^{[LSF]} } \quad .
\end{equation}
Figure~\ref{fig:lineShape} illustrates $\omega$ for Voigt profiles, as a function of the FWHM of the Gaussian and Lorentz contributions. On the horizontal axis is the FWHM of the Gaussian, $F_G$, on the vertical axis is the FWHM of the Lorentz profile, $F_L$. The FWHM of the resulting Voigt profile, $F_V$ is in good approximation \citep{Olivero1977}
\begin{equation}
F_V = 0.5356 \cdot F_L + \sqrt{0.2166 \cdot F_L^2 + F_G^2} \quad .
\end{equation}
The pseudo-LSF used here is based on the optical LSF for \gaia, blurred with a Gaussian with a standard deviation of 0.64, which is the LSF model used in Sect.~\ref{sec:Examples} of this work. The LSF changes with wavelength, and for this example we use the wavelength of H$\beta$. For lines with FWHM below about 20~nm, the increase in $\omega$ is only a few ten per cent, showing the intuitive result that low resolution spectra are thus not particularly sensitive to line widths.

\subsection{Continuum model for broad lines \label{sec:continuumBroad}}

In the case of broad lines, estimating the continuum underlying a spectral line requires some form of interpolation. We would like to do the interpolation on internally calibrated spectra, where we detect a line, but at the same time we would like to ensure that the interpolation on the internally calibrated spectrum corresponds to a smooth interpolation on the externally calibrated spectrum as well. To achieve this, we may adapt the approach already discussed in Sect.~\ref{sec:inversion} and choose smooth interpolation functions in the external regime, and compute the observational spectra corresponding to these interpolation functions. We can then use these artificial observational spectra of the interpolation functions to find an interpolation over some pseudo-wavelength interval for an XP spectrum. For the interpolation functions, we may choose again the first $M$ Legendre polynomials over a wavelength interval that essentially covers the entire range of non-zero response in BP and RP, respectively.\par
With the interpolation functions at hand, we have to decide about the interpolation interval $[u_{\rm lower},u_{\rm upper}]$ under the line. The choice of this interval is crucial for a good estimate of the continuum. Choosing this interval too narrow will result in an under-subtraction of the continuum, while choosing it too wide is likely to make the interpolation unreliable. We tie the interval to the widths of the line, as expressed by the distance between the two inflection points on either side of the local extremum, $u_{i,2} - u_{i,1}$, as discussed in the previous subsection. We choose the interpolation interval symmetric around the local extremum, with a width corresponding to a factor $x$ times $\Delta$. Experimenting with simulated spectra with different Black Body continua and Voigt profiles for the lines, we found a value around $x \approx 5.5$ is a good choice.\par
For the interpolation, we again have to impose conditions on the function at the boundaries of the interpolation interval and different points of the function itself on either side if the interval. So we select a small number of $m$ points of either side of the interval, separated by a value $\delta$, such that we obtain a sequence of $2m+2$ points ${\bf u} \coloneqq u-m\delta,u-(m-1)\delta,\ldots,u_{\rm lower},u_{\rm upper},u+\delta,\ldots,u+m\delta$. We might not force an exact solution for the interpolation, but chose $2m+2> M$, and do a regression, as it reduces the noise. The noise on $f({\bf u})$ however is likely to be correlated, with a covariance matrix $\Sigma_f = {\bf \Phi}^{\mathsf T}({\bf u}) \, \Sigma_c \, {\bf \Phi}({\bf u})$. We thus have to determine the coefficient vector for the interpolation functions, $\bf b$, such that it minimised the squared Mahalanobis distance as a generalised least squares problem, as already discussed in Sect.~\ref{sec:inversion}. The product ${\bf L} {\bf b}$ provides then the coefficients for the continuum approximation on the interval $[u_{\rm lower},u_{\rm upper}]$.

\subsection{Equivalent widths for broad lines \label{sec:WforBroadLines}}

Having separated the line and the continuum, the computation of the equivalent width requires the inversion of both components of the spectrum. The inversion of the continuum is trivial, as the coefficients $\bf b$ have been constructed from the observational spectra corresponding to Legendre polynomials in the regime of the absolute calibration. However, the inversion of the line signal is intrinsically an ill-posed problem and any inversion might be subject to significant systematic errors. We therefore consider a simplification and compute an estimate for the equivalent width without inversion of the internally calibrated spectrum, i.e. using the approximation
\begin{equation}
W \approx \int\limits_{u_1}^{u_2} \frac{f(u)-p({\bf b},u)}{p({\bf b},u)} \, \frac{{\rm d}\lambda}{{\rm d}u} \, {\rm d}u \quad ,
\end{equation}
with $p({\bf b},u)$ the continuum approximation. The approximate computation of $W$ from the internally calibrated spectrum instead of the SPD has been tested using simulations for narrow lines and Voigt profiles with different underlying continua. The resulting systematic errors were in the order of some per cent. In all test cases, the systematic errors from continuum interpolation were at least three times larger than the error resulting from the approximate computation. This approximation might for general broad lines be sufficiently good.\par
Example computations for the determination of $W$ with this approach are shown in Fig.~\ref{fig:lineInterpolationExample}. The simulated case is analogous to the case considered for narrow lines, a Black Body continuum for $2 \times 10^4$~K and an H$\beta$ line are used, for $G=12^{\rm m}$ and 10 transits, and $5 \times 10^5$ random realisations of the BP spectrum with randomly selected $W$. Panel B shows the distribution of the relative difference between the computed and the true equivalent width as a function of true equivalent width. The performance of the broad line approach is similar to the use of the narrow line approach, shown in panel A. The random noise is slightly larger, though.\par
Panel C of the same figure shows the distribution for an H$\beta$ line with a Gaussian profile and a full width at half maximum (FWHM) of 15~nm. In this case, the Legendre interpolation was done without taking correlations into account. A systematic error in $W$ can be seen, resulting mainly from a systematic underestimation of the continuum in the fit. Panel D shows the same case with taking correlations into account. The random noise is suppressed as compared to the case without including correlations, but the systematic errors from the continuum interpolation remain. Panel F of Fig.~\ref{fig:lineInterpolationExample} shows the results for a Voigt profile for the H$\beta$ line, with a FWHM of 15~nm and equal FWHM for the Gaussian and Lorentz contributions to the Voigt profile. The systematic errors from the continuum interpolation increase as compared to a Gaussian line profile. Panel H shows the distribution of relative difference in $W$ for a Lorentz profile with a FWHM of 15~nm. Here, the systematic errors from continuum interpolation increase even more. The reason for the increasing systematic errors from Gaussian to Voigt to Lorentz profile are the increasingly flat wings of the spectral line, which make a choice of a suitable interpolation interval for the continuum increasingly difficult. As a result, the fit is likely to over- or under-subtract the line from the continuum. This problem can be strongly reduced if the shape of the line can be assumed known a-priori, a case which we discuss in the following section.

\subsection{Broad lines with known line shapes \label{sec:broadKnown}}

In cases where the position and shape of a broad line can be assumed to be known a-priori, the analysis of the line in \gaia~DR3 XP spectra simplifies. In this case, the interpolation step for the continuum can be replaced by an approach more similar to the case of narrow lines. Assuming a line profile $p(\lambda)$, normalised such that its integral is unity, we compute the observational spectrum corresponding to this profile,
\begin{equation}
P(u) \coloneqq \int\limits_{0}^{\infty} L(u,\lambda) \, R(\lambda) \, p(\lambda) \, {\rm d}\lambda \quad. 
\end{equation}
Analogous to the approach in Sect.~\ref{sec:continuumNarrow} we then use the second derivative at the line position to determine the scaling factor $\alpha$ that describes the line strength:
\begin{equation}
\alpha = \frac{f^{(2)}(u_L)}{P^{(2)}(u_L)} \quad .
\end{equation}
The approximation for the continuum spectrum is
\begin{equation}
f_c(u) \approx f(u) - \alpha \, P(u) \quad .
\end{equation}
As the line profile is assumed to be known, no numerical inversion of the profile is required. An inversion of the continuum over the pseudo-wavelength range covered by the line is however required. As the continuum can be assumed smooth, the inversion for $t_c(u)$ using the Taylor approximation as lined out in Sect.~\ref{sec:inversion} might allow for a good smooth approximation on a grid of points in the interval of interest. With knowledge of the response function, we can thus obtain an estimate for $s_c(\lambda)$, and compute the equivalent with by numerical integration of the expression
\begin{equation}
W = \alpha \, \int\limits_{\lambda_1}^{\lambda_2} \frac{p(\lambda)}{s_c(\lambda)} \, {\rm d} \lambda \quad .
\end{equation}
The inversion of the continuum point-wise is increasing the computational effort, but in exchange it is possible to remove systematic errors resulting from the interpolation of the continuum. Illustrations are shown in panel E, G, and I of Fig.~\ref{fig:lineInterpolationExample}, for a Gaussian, a Voigt, and a Lorentz line profile, respectively. The equivalent width of an H$\beta$ line with a Black Body continuum of $2\times 10^4$ K and a FWHM of 15~nm is computed assuming the line position and profile known. The distribution of the relative difference between the computed and true $W$ is shown for $5 \times 10^5$ random generations for $G =12^{\rm m}$ and 10 transits. The systematic error in $W$ that results when interpolating the continuum without using a-priori knowledge about the line, a case shown in panels D, F, and H of the same figure, is strongly reduced. The reduction in systematic error is strongest for a Gaussian line profile, and lowest for the Lorentz profile. The differences in the remaining systematic error mainly arise from the interval of integration. From Gaussian and Voigt to Lorentz profile, the wings of the line become wider, and the systematic errors from an integration over a fixed interval become larger. The reduction in systematic error however comes with an increase in the random error.\par
If the position and shape of the line is known a-priori, it is possible to compute an upper limit for a non-detected line, following the approach discussed for narrow lines in Sect.~\ref{sec:upperLimits}, or treat the problem of blended lines analogous to the case of narrow lines described in Sect.~\ref{sec:blended}.

\section{Lines as extrema in higher derivatives\label{sec:higherOrder}}
As we have seen in Sect.~\ref{sec:systematicErrorInPosition}, no local extremum in the observational spectrum is formed by a line if the function $\zeta(u)/W$ is not intersecting with the function $L^{(1)}(u,u_L)$, the first derivative of the LSF or pseudo-LSF. As can be seen from Eq.~(\ref{eq:zeta}), large values for $\zeta(u)/W$ are a result of a too large ratio between the slope of the continuum spectrum, $f^{(1)}_c(u)$, and the line's equivalent width $W$. Thus, a large slope and a weak line, i.e. a line with a small absolute value of $W$, might prevent the formation of a local extremum. However, a line might still be able to produce a local extremum in higher derivatives. We are therefore considering higher derivatives of Eq.~(\ref{eq:linePosition}), stating the condition for a local extremum.\par
\begin{figure}
   \centering
   \includegraphics[width=0.49\textwidth]{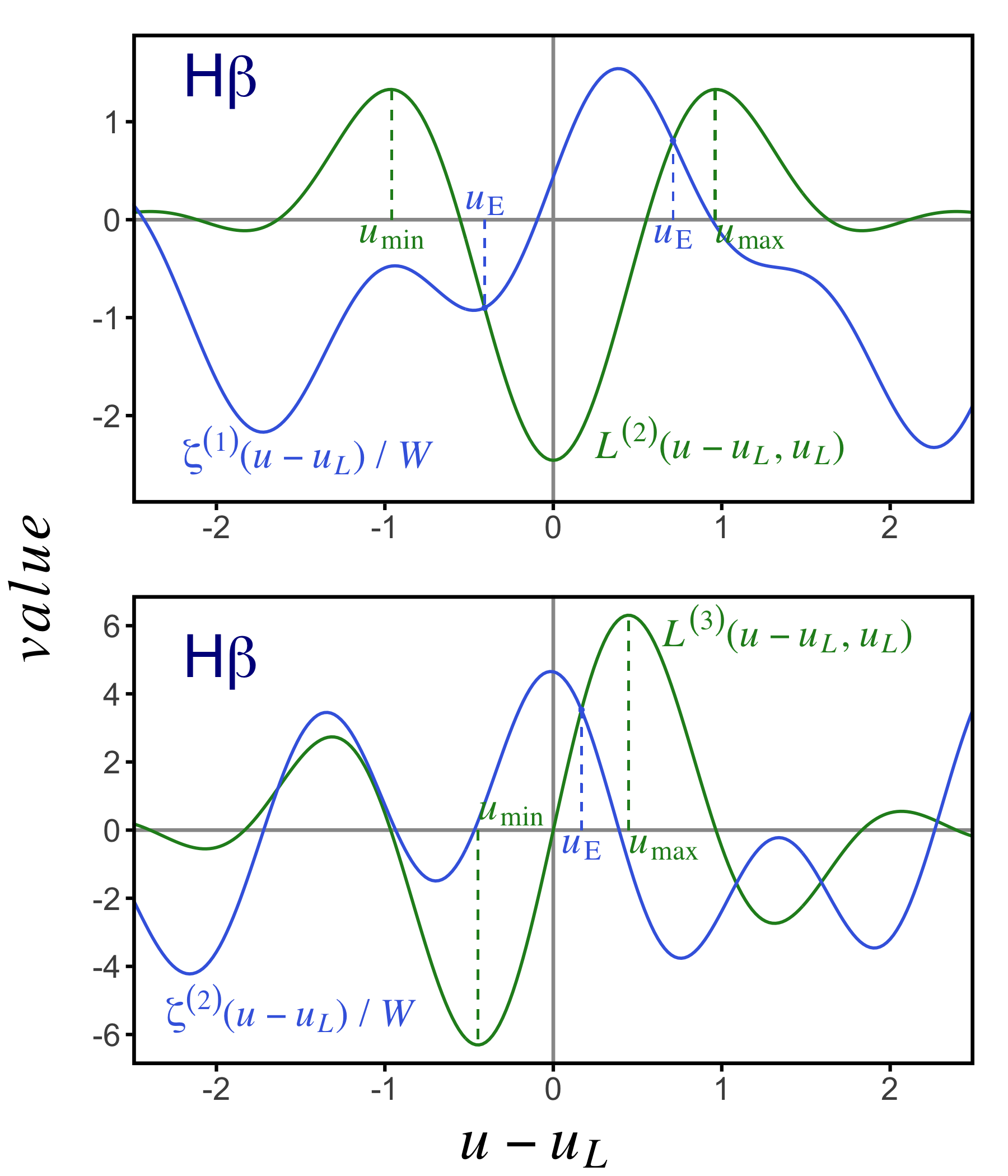}
   \caption{Illustration of the positions of local extrema in the first derivative (top panel) and second derivative (bottom panel) for a simulated H$\beta$ line with a 6000K Black Body continuum. For more explanation see text.}
              \label{fig:systematicShiftHigherDerivative}
    \end{figure}

The function $L^{(2)}(u,u_L)$ has no root at $u_L$, and a line might show as two local extrema in $f^{(1)}(u)$ on both sides of $u_L$. These two points correspond to inflection points of the observational spectrum. This situation is illustrated in the top panel of Fig.~\ref{fig:systematicShiftHigherDerivative}, for a simulated H$\beta$ line with a Black Body continuum of 6000K. The shapes of the higher derivatives of the pseudo-LSF and $\zeta(u)/W$ become more complex, and the positions of $u_E$ might vary strongly. A good indicator for inflection points being related to a line is whether the values of $u_E$ are within the first local extrema of $L^{(2)}(u_L)$, indicated by $u_{min}$ and $u_{max}$ in Fig.~\ref{fig:systematicShiftHigherDerivative}.\par
\begin{figure}
   \centering
   \includegraphics[width=0.49\textwidth]{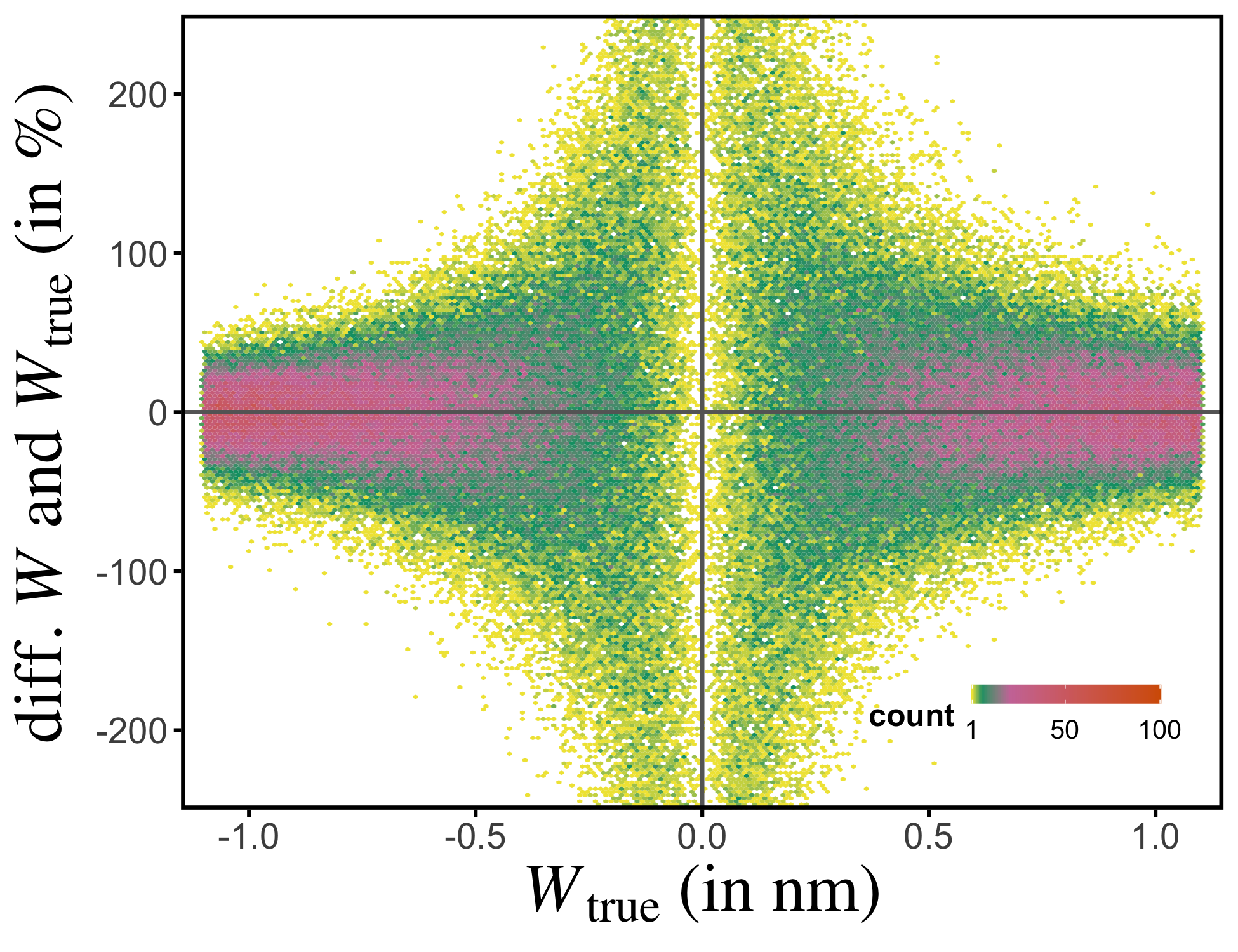}
   \caption{Distribution of the difference between true and derived equivalent width for simulated weak and narrow H$\beta$ lines. For details see Sect.~\ref{sec:higherOrder}.}
              \label{fig:lineInterpolationExample2}
    \end{figure}

The function $L^{(3)}(u,u_L)$ has a root at $u_L$, as the curvature of the pseudo-LSF is maximal at its centre. We might therefore consider the case of local extrema in $f^{(2)}(u)$. This case is illustrated in the bottom panel of Fig.~\ref{fig:systematicShiftHigherDerivative}. For the determination of the equivalent width of a narrow line we might therefore use the approach of Eq.~(\ref{eq:W1}), with the derivatives increased by two,
\begin{equation}
W = \frac{f^{(4)}(u_E) }{L^{(4)}(u_E,u_L)} \frac{1}{t_c(u_L)}\quad .
\end{equation}
As an example, we consider again a simulated H$\beta$ line in BP, with an underlying Black Body continuum for a temperature of 6000 K. This simple test case does not result in a local extremum for $-1.1\text{ nm} \le W \le 1.1 \text{ nm}$. We derive the equivalent width from the extremum in the second derivative of $f(u)$. The resulting distribution of differences between the true and the computed $W$, as a function of true $W$, is shown in Fig.~\ref{fig:lineInterpolationExample2}. This figure shows the result from $2 \times 10^{5}$ random realisations of an object, with $G = 12^{\rm m}$ and 10 transits. Unbiased recovery of the equivalent width is possible, with some tens of percent uncertainty, also from the second derivative of a spectrum. As expected for faint lines and the use of higher derivatives, the random noise is considerably larger than for the case of deriving equivalent widths from local extrema. However, only if $|W|$ falls below about 0.5~nm, the noise becomes so large that no meaningful interpretation of the spectrum around the H$\beta$ line becomes possible in the simulated case.

\section{Example applications to \textit{Gaia}~DR3 spectra \label{sec:Examples}}

In the following we apply the methods derived in the previous sections to some example cases among the \gaia~DR3 low resolution spectra. These examples are not selected for their scientific output, but serve solely for illustration of the computational method, and for illustrating its potential. To cover a wide range of applications, the examples in this section include the analysis of lines in the zero and second derivative, absorption and emission lines, narrow and broad lines, and blended and non blended cases, for hydrogen Balmer lines, He I lines, and broad interstellar absorption bands.\par
For the instrument model, we use the dispersion functions and response curves from the configuration files in the \gaia~software tool GaiaXPy\footnote{\url{https://gaia-dpci.github.io/GaiaXPy-website/}, see \cite{DeAngeli2022}}. For easy computation, we represent the tabulated dispersion functions by spline functions. The configuration files contain no model for the (pseudo-)LSF. To construct a simple substitute, we measure the widths $\omega$ of H$\alpha$ and H$\beta$ lines in a large set of sources known for strong hydrogen absorption and emission with respect to the optical LSF of \gaia, broadened with pixel sampling and TDI effects (cf. \citealt{Weiler2020}). We then applied a Gaussian blur to this LSF until the width of the resulting blurred LSF matched the observed widths for H$\alpha$ and H$\beta$. The standard deviation of the Gaussian blur applied is 0.64, independent of wavelength. From this simple LSF model, we compute the pseudo-LSF by development in Hermite functions. In the following we will show that this simplistic approach is already sufficient to obtain good results. However, the results presented should be understood as preliminary, as they are likely to improve with the use of a realistic full instrument model when available in the future.\par
The analysis of the H$\alpha$ line of a large set of sources with strong H$\alpha$ emission showed a systematic difference between the pseudo-wavelength at which the RP dispersion function predicts this line and the true line position by about 0.3 in pseudo-wavelength. We corrected for this systematic effect by artificially shifting the value of $u_L$ for H$\alpha$ by +0.3 with respect to $u_L$ predicted by the dispersion relation used. A similar effect seems to apply at long wavelengths for the BP instrument, and we apply a correction of +0.35 for the He I line at 587.7nm, too.\\

\subsection{Example: Balmer emission and absorption lines \label{sec:Balmer}}

\begin{figure}
   \centering
   \includegraphics[width=0.49\textwidth]{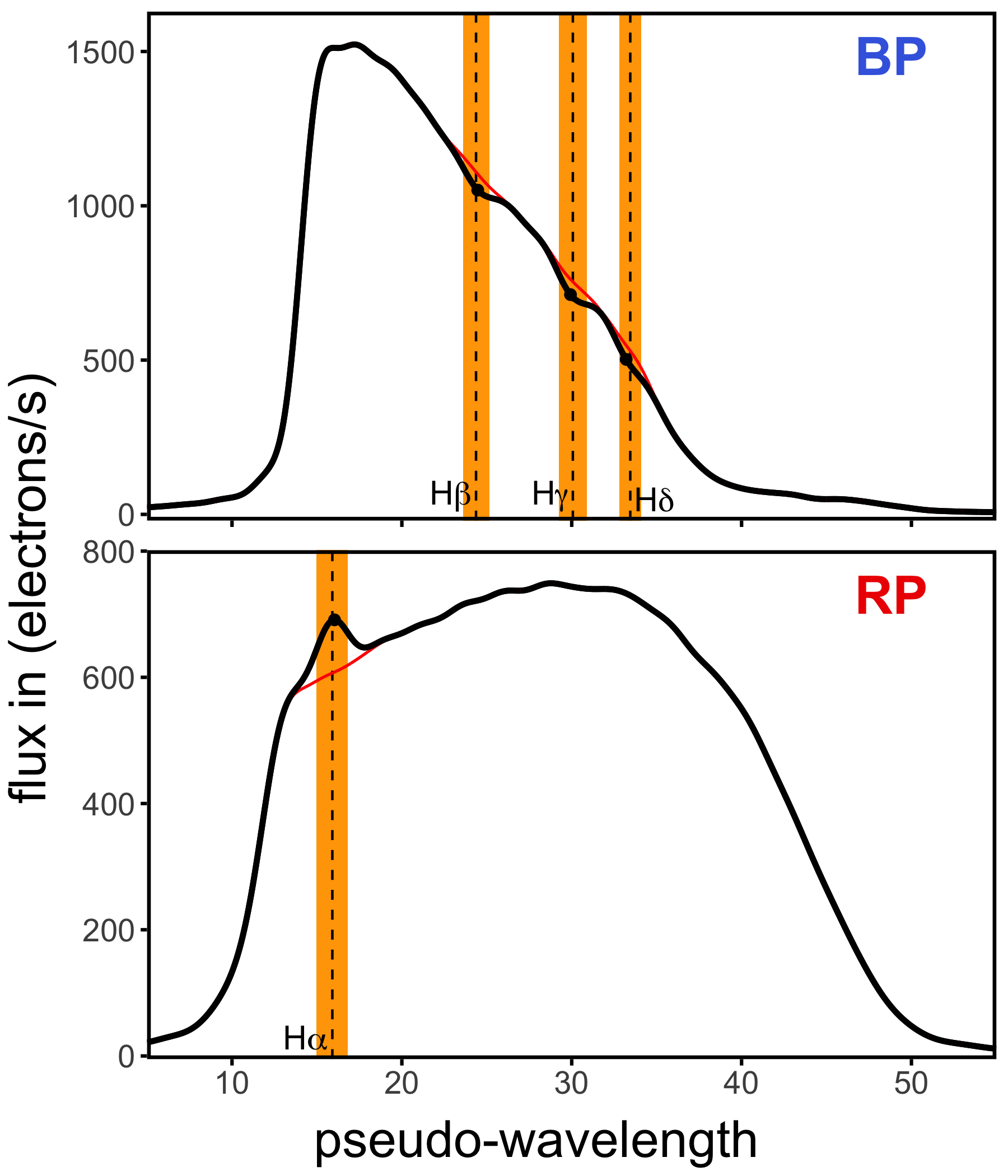}
   \caption{Example for the analysis of H$\beta$ to H$\delta$ in BP (top panel) and H$\alpha$ in RP (bottom panel) for source {\tt Gaia DR3 505171240162594560}. The nominal line positions are indicated by dashed lines, the shaded regions indicate the range in pseudo-wavelength within which a local extremum is considered in agreement with being a hydrogen Balmer line. For BP, the wavelength increase from right to left (with decreasing pseudo-wavelength), for RP from left to right (with increasing pseudo-wavelength). The black curves show the XP spectra, the red curves the continuum model. The black dots indicate the positions of local extrema in the second derivative (H$\beta$ to H$\delta$) and zero derivative (H$\alpha$).}
              \label{fig:balmer}
    \end{figure}

As a first example we consider the star {\tt Gaia DR3 505171240162594560}, IRAS 02058+5719, an object with $G \approx 14.187^{\rm m}$ and $G_{BP}-G_{RP} \approx 0.495^{\rm m}$. This object has been selected from the IGAPS data set \citep{Monguio2020}, where H$\alpha$ emission is reported. The analysis of this object is illustrated with the BP and RP spectra as a function of pseudo-wavelength in Fig.~\ref{fig:balmer}. The expected position of the H$\alpha$ to H$\delta$ lines are indicated by dashed lines in this figure. The intervals $[u_{\rm min},u_{\rm max}]$, within which an extremum is consistent with resulting from the Balmer lines, are highlighted. The intervals are based on the two inflection points of the LSF on both sides of the maximum position of the LSF, respectively, as discussed in Sect.~\ref{sec:systematicErrorInPosition}. For this star, we find a local maximum in RP consistent with H$\alpha$, and three local extrema in the second derivative, consistent with H$\beta$, H$\gamma$, and H$\delta$. All four local extrema have $p$-values larger than 0.999 and it is thus very unlikely that these are noise patterns. The red lines show the continuum model for narrow lines according to Eq.~(\ref{eq:continuum}). We use Eq.~(\ref{eq:W1}) for the computation of the equivalent widths of the lines, using the inversion according to Eq.~(\ref{eq:localInversion}). The results for the four Balmer lines in the XP spectra of this source are summarised in Table~\ref{tab:WD}. The difference between the wavelength corresponding to the local extrema, $\lambda_e$, and the wavelengths of the corresponding Balmer lines is well within the range of expected systematic deviations. The estimated line widths, $\omega$, are in a good agreement with the assumption of narrow lines. Only the H$\gamma$ line shows a deviation above a 1-$\sigma$ level, but mildly so.

\begin{table*}
\center
\renewcommand\arraystretch{1.2}
\caption{Line, XP instrument (BP or RP), position in pseudo-wavelength ($u_E$) and shift with respect to expected line position ($\Delta u$), wavelength ($\lambda_e$) for the local extrema, line width parameter $\omega$, estimated equivalent width ($W$), and $p$-value for the lines in the XP spectra of individual sources analysed in this work. \label{tab:WD}}
\begin{tabular}{c c c c c c c c} \hline
Line             & XP & $u_E$ & $\Delta u$ & $\lambda_e$ (nm) & $\omega$ & $W$ (nm) & $p$ \\ \hline
\multicolumn{8}{c}{{\tt Gaia DR3 505171240162594560} (IRAS 02058+5719)}\\ \hline
H$\alpha$   & RP &  16.034 $\pm$ 0.037 & \phantom{$-$}0.127 & 657.34 $\pm$ 0.25 & 0.984 $\pm$ 0.030 & \phantom{$-$}2.35 $\pm$ 0.19 &1.000 \\
H$\beta$     & BP &  24.466 $\pm$ 0.093 & \phantom{$-$}0.100 & 485.20 $\pm$ 1.01 & 1.062 $\pm$ 0.106 & $-$1.13 $\pm$ 0.32 & 1.000 \\
H$\gamma$ & BP &  29.927 $\pm$ 0.073 & $-$0.139 & 435.25 $\pm$ 0.57 & 1.229 $\pm$ 0.154 &  $-$0.98 $\pm$ 0.26 & 1.000 \\
H$\delta$    & BP&  33.206 $\pm$ 0.057 & $-$0.242 & 411.86 $\pm$ 0.37 & 0.972 $\pm$ 0.036 & $-$1.47 $\pm$ 0.28 & 1.000 \\ \hline
\multicolumn{8}{c}{{\tt Gaia DR3 426306363477869696} (HT Cas)}\\ \hline
He I (706.7nm) & RP &  21.841 $\pm$ 0.333 & $-$0.458 & 702.95 $\pm$ 2.70 & 0.76 $\pm$ 0.12  & \phantom{$-$}0.55 $\pm$ 2.31 & 0.420 \\
H$\alpha$ & RP &  16.005 $\pm$ 0.048 & \phantom{$-$}0.098 & 657.14 $\pm$ 0.33 & 1.004 $\pm$ 0.050 & \phantom{$-$}12.09 $\pm$ 1.89 & 1.000 \\
He I (587.7nm) & BP &  16.932 $\pm$ 0.550 & $-$0.621 & 586.25 $\pm$ 10.10 & 0.65 $\pm$ 0.25 & \phantom{$-$}1.51 $\pm$ 2.26 & 0.270 \\
H$\beta$ & BP &  24.319 $\pm$ 0.086 & $-$0.048 & 486.80 $\pm$ 0.93 & 0.961 $\pm$ 0.078 & \phantom{$-$}6.93 $\pm$ 1.61 & 1.000 \\
He I (447.3nm) & BP &  27.816 $\pm$ 0.274 & $-$0.624 & 452.67 $\pm$ 2.40 & 0.764 $\pm$ 0.079 & \phantom{$-$}2.17 $\pm$ 1.40 & 0.700 \\
H$\gamma$ & BP &  29.706 $\pm$ 0.161 & $-$0.360 & 436.98 $\pm$ 1.27 & 0.858 $\pm$ 0.077 & \phantom{$-$}5.86 $\pm$ 1.99 & 0.969 \\
H$\delta$ & BP &  33.067 $\pm$ 0.169 & $-$0.380 & 412.77 $\pm$ 1.12 & 0.859 $\pm$ 0.091 &\phantom{$-$} 5.65 $\pm$ 2.01 & 0.957 \\
H$\epsilon$ & BP &  35.768 $\pm$ 0.126 & \phantom{$-$}0.174 & 396.12 $\pm$ 0.73 & 0.833 $\pm$ 0.048 & \phantom{$-$}3.30 $\pm$ 0.89 & 0.978 \\ \hline
\multicolumn{8}{c}{{\tt Gaia DR3 2067888218857234304} (BD +41 3807, ALS 11438)}\\ \hline
770nm DIB & RP &  28.968 $\pm$ 0.105 & $-$0.303 & 766.92 $\pm$ 1.04 & 1.15 $\pm$ 0.16 & $-$1.38 $\pm$ 0.55 & 1.000 \\ \hline
\end{tabular}
\end{table*}

\subsection{Example: H$\alpha$ absorption for CALSPEC stars}

In order to test the accuracy of the equivalent widths derived from \gaia~XP spectra with the method presented in this work, we compare the values of $W$ with values derived from higher resolution spectra of the same sources. As a set of stars with high quality spectra with higher resolution we use the CALSPEC data \citep{Bohlin2014b,Bohlin2020}, and we consider the H$\alpha$ line. In the visible wavelength range CALSPEC spectra are obtained with the Hubble Space Telescope / Space Telescope Imaging Spectrograph (HST/STIS) and the resolving power at the H$\alpha$ line is approximately 673. The data set provides 44 stars with XP spectra available in DR3 and a detection of H$\alpha$ in the RP spectrum with a $p$-value larger than 0.8.\par
In the CALSPEC spectrum of each star a third degree polynomial was fitted to continuum intervals on both sides of the line and then used as an approximation of the continuum. With the continuum approximation, the FWHM of the line is derived and the equivalent width computed by applying Eq.~(\ref{eq:defW}).  The 44 stars cover a wide range of parameters, with equivalent widths between $-0.2$ nm and $-4.4$ nm,  FWHM of the lines ranging from 1.5 nm to 7.8 nm, $G$ band magnitudes from $14.7^{\rm m}$ to 4.3$^{\rm m}$, and $G_{\rm BP} - G_{\rm RP}$ colours ranging from $-0.54$ to $0.91$.\par
\begin{figure}
   \centering
   \includegraphics[width=0.45\textwidth]{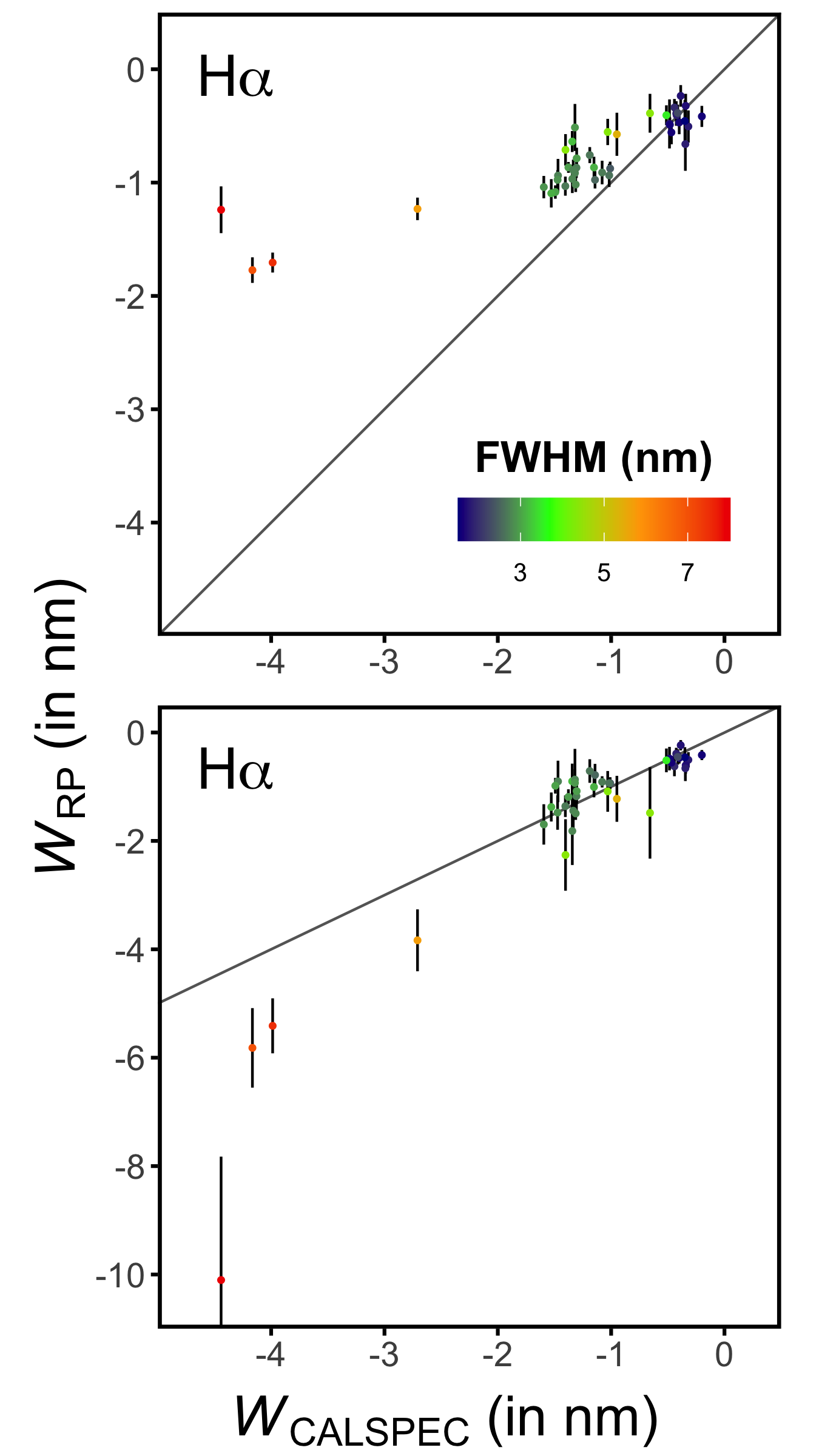}
   \caption{Comparison of the equivalent widths of the H$\alpha$ line for CALSPEC stars, derived from \gaia~RP spectra, $W_{\rm RP}$, and from CALSPEC spectra, $W_{\rm CALSPEC}$. Top panel: Results obtained with the narrow line approximation. Bottom panel: Results for assuming a Lorentz line profile. The colour scale gives the FWHM of the lines, as measured in CALSPEC spectra, and is the same for both panels.}
              \label{fig:calspec}
    \end{figure}

A comparison between the equivalent widths derived from RP spectra and from the CALSPEC spectra is shown in Fig.~\ref{fig:calspec}. The upper panel of this figure shows the equivalent widths derived with the narrow line approximation as described in Sect.~\ref{sec:WforNarrowLines}. While this approximation is providing a good agreement with the reference values from CALSPEC spectra for weak absorption lines, systematic deviations are obvious for stronger absorption lines. Already for  equivalent widths around $-1.5$ nm the systematic error is clearly visible.\par
The bottom panel of Fig.~\ref{fig:calspec} shows the comparison for equivalent widths computed by assuming a Lorentz line shape for the H$\alpha$ line for all lines with a value of $\omega$ larger than one, and using the approach described in Sect.~\ref{sec:broadKnown}. The width of the Lorentz profile was chosen such that the value of $\omega$ derived for each source from the RP spectrum is met by the model Lorentz line. With this approach, the systematic error is strongly reduced, while the random error, which has been estimated by using bootstrapping, increased. This is consistent with the results obtained for simulated spectra in Sect.~\ref{sec:broadKnown}. Only the strongest and broadest lines in Fig.~\ref{fig:calspec} might be affected by systematics. However, besides the systematic errors resulting from the computational approach used, also the simplistic instrument model employed in this work might contribute to systematic errors for the broad and strong absorption lines.

\subsection{Example: H$\alpha$ and H$\beta$ absorption in Open Clusters \label{sec:Balmer}}

\begin{figure}
   \centering
   \includegraphics[width=0.49\textwidth]{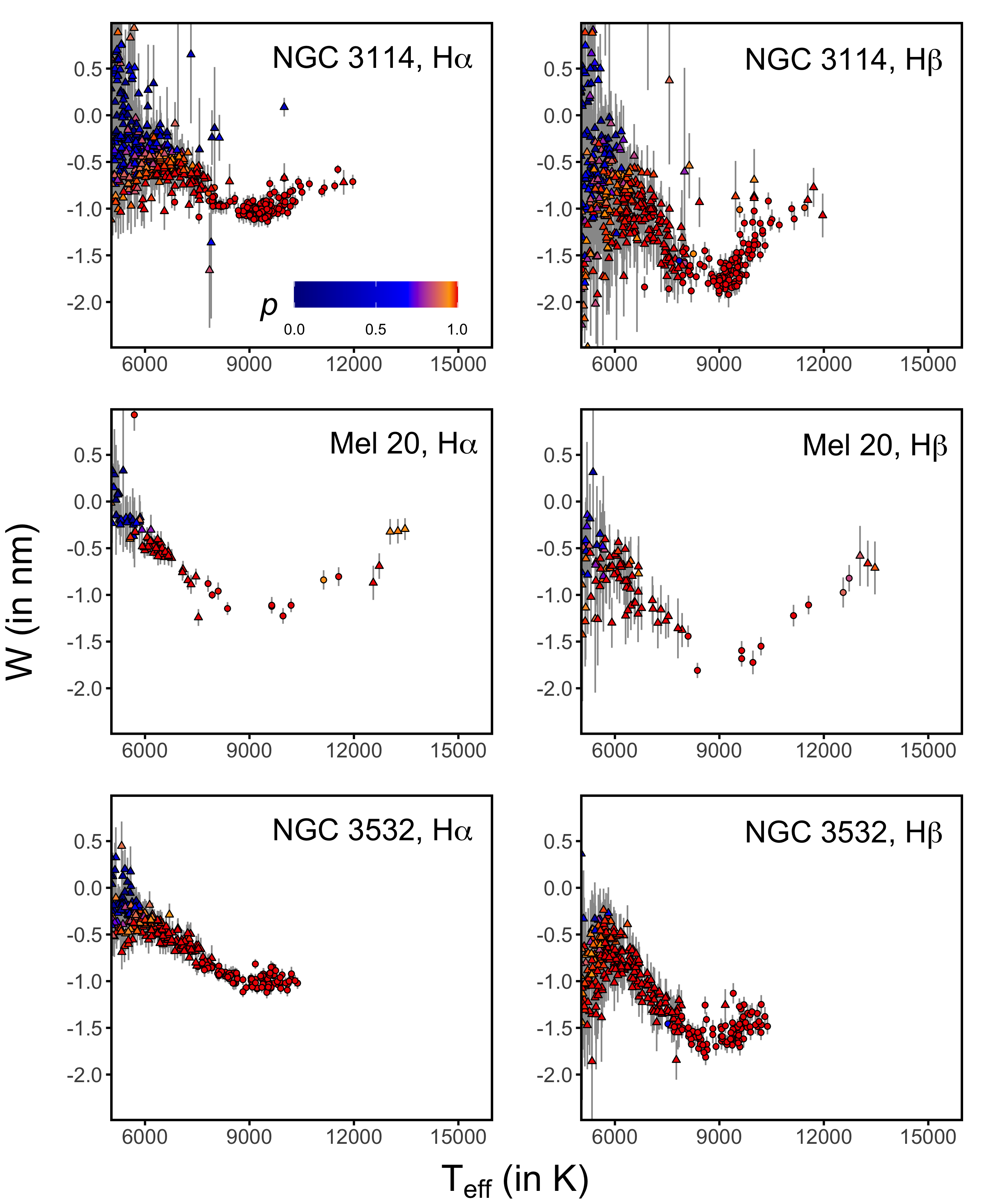}
   \caption{H$\alpha$ and H$\beta$ equivalent widths against the effective temperature for the three open clusters NGC 3114, Mel 20, and NGC 3532. The colour scale shows the $p-$value and is common to all panels. Circles correspond to extrema in the zero derivative, triangles to extrema in the second derivative.}
              \label{fig:oc}
    \end{figure}

As an example for the application of the method presented in this work to a larger set of sources, we select open clusters, and determine the H$\alpha$ and H$\beta$ equivalent widths. For the test we selected the three open clusters NGC 3114, Mel 20 ($\alpha$ Persei), and NGC 3532, the cluster members taken from \cite{Cantat-Gaudin2020}. For the test we applied strict quality filters to the data, including only sources with fractions of blended and contaminated transits below 0.1, a small renormalised unit weight error below 1.4, a small number of multi-peaked transits or transits with complex windows, and an absolute value of the corrected flux excess below $1.5 \sigma_{C^\ast}$ according to \cite{Riello2021}. This filtering resulted in 851, 552, and 820 sources for the three open clusters, respectively. Figure~\ref{fig:oc} shows the equivalent widths as a function of effective temperature for the three open clusters. The effective temperatures are the field {\tt teff\_gspphot} from the DR3 gaia\_source and astrophysical parameters table \citep{Creevey2022,Fouesneau2022}. The $p-$value of the hydrogen lines is shown as a colour scale, and lines which are local extrema in the XP spectra are shown as circles, the ones which are extrema in the second derivative of the XP spectrum as triangles. The generally expected behaviour can be seen in the results for all three clusters, and for H$\alpha$ and H$\beta$. Weaker lines tend to be detected in the second derivative, and the statistical significance of the lines, expressed by the $p-$value, increases as the lines are getting stronger. Also the errors on the equivalent widths tend to become smaller if the line strength increases. Also as expected, a maximum hydrogen absorption around an effective temperature of 9000~K can be seen in all three clusters and for H$\alpha$ and H$\beta$.\par
The open cluster NGC~3114 shows one prominent outlier from the general trend of $H\alpha$ with effective temperature, which is {\tt Gaia DR3 5256686653442846336} (HD~87471). This star with an effective temperature of about $10^4$~K has no significant H$\alpha$ line, with $W = (0.087 \pm 0.100)$~nm, and a $p$-value of 0.414. The H$\beta$ absorption for this star is also somewhat lower for this effective temperature, with $W = (-0.87 \pm 0.23)$~nm, with a $p$-value of $p = 0.9999$. No spectral peculiarities have been reported for this star. It is classified as a sub-giant \citep{Houk1975}, and this classification is consistent with the fact that the star is approximately 1.5 magnitudes brighter than the main sequence stars of comparable colour in the same cluster. A lower hydrogen absorption as compared to the other stars in the cluster might thus result at least partially from a lower surface gravity of a sub-giant.

\subsection{Example: Blended narrow lines}

\begin{figure}
   \centering
   \includegraphics[width=0.49\textwidth]{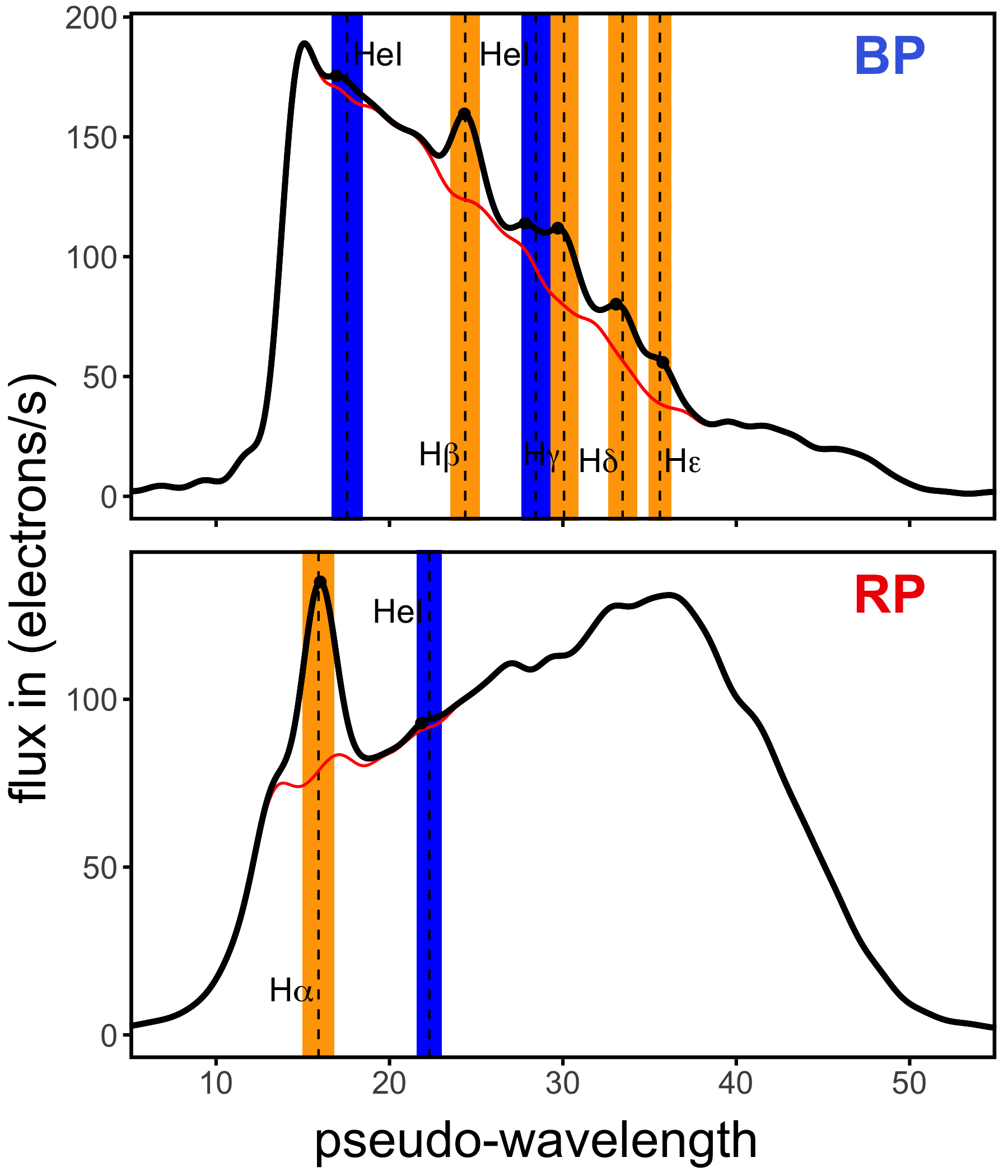}
   \caption{Analysis of hydrogen Balmer and He I lines for {\tt Gaia DR3 426306363477869696}. As in Fig.~\ref{fig:balmer}, the wavelengths increase from right to left for BP (upper panel), and from left to right for RP (bottom panel).}
              \label{fig:HHe}
    \end{figure}

As an example for the treatment of blended narrow lines we chose the cataclysmic variable HT Cassiopeiae, {\tt Gaia DR3 426306363477869696}. This star, with a $G \approx 16.313$\m~ and a $G_{BP} - G_{RP} \approx 0.664$\m~, is known to show strong emissions from hydrogen and neutral helium \citep{Marsh1990}. We therefore study the hydrogen Balmer lines and three strong He I lines within the wavelength range covered by BP and RP, at wavelengths of 447.3nm, 587.7nm, and 706.7nm, which are the vacuum Ritz wavelengths from the National Institute of Standards and Technology Atomic Spectra Data Base \citep{NIST_ASD}. The XP spectra for this source as a function of pseudo-wavelength are shown in Fig.~\ref{fig:HHe} for illustration. The derived line parameters are presented in Table~\ref{tab:WD}. All hydrogen Balmer lines for H$\alpha$ to H$\epsilon$ are clearly detected in emission, with $p-$values above 0.95 in all cases. There are local extrema in the XP spectra or its second derivative that are consistent with being He I emissions. However, the statistical significance of these extrema is rather low. The $p-$values for the lines at 706.7nm and 587.7nm are 0.420 and 0.270, respectively, and the extrema are thus well in agreement with being noise. The extremum consistent with the 447.3nm He I line has a somewhat larger $p-$value of 0.700, and might be a tentative detection of He I.\par
The lines of H$\beta$, He I at 447.3nm, and H$\gamma$ to H$\epsilon$ are blended, and the equivalent widths of these lines are computed using the approach presented in Sect. \ref{sec:blended}. With five blended lines, the continuum model is constructed over a wide range of pseudo-wavelengths, shown as the red line in Fig.~\ref{fig:HHe}. The continuum model has a wavy structure, partly because it is affected by the noise. Additionally, the assumptions on the LSF made in this work are simplified, and in particular the outer parts of the pseudo-LSF are affected by these simplifications. When combining the outer parts of the pseudo-LSF in case of blended lines, these simplifications might therefore additionally affect the resulting solution. An improved LSF model might therefore allow for improvements in the future.

\subsection{Example: Broad interstellar absorption line}

\begin{figure}
   \centering
   \includegraphics[width=0.49\textwidth]{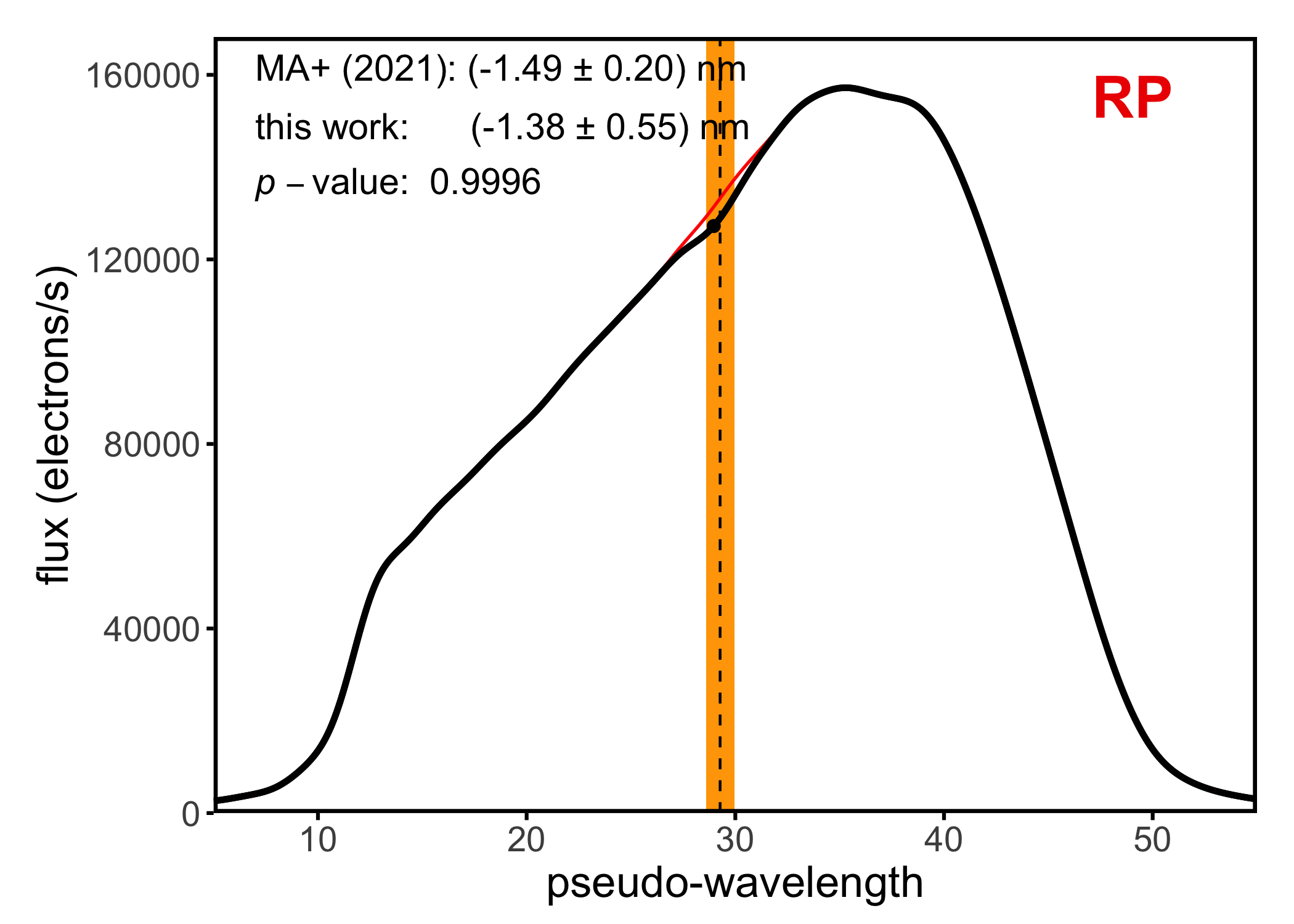}
   \caption{Example for the analysis of  the broad interstellar band at 770nm for source {\tt Gaia DR3 2067888218857234304}. The nominal band position is indicated by dashed lines, the shaded regions indicate the range in pseudo-wavelength within which a local extremum is considered in agreement with being the band. The black curve shows the XP spectra, the red curve the continuum model. The black dot indicates the positions of local extrema in the second derivative.}
              \label{fig:dib}
    \end{figure}

As an example for a broad line we consider the interstellar band at approximately 770 nm in RP. \cite{Maiz2021} discovered this band in reddened OB stars and provides a number of stars with equivalent widths for this band. The band shape is Gaussian, with a central wavelength of 769.9 nm and a FWHM of 17.66 nm. We use this information on the line shape to apply the approach discussed in Sect. \ref{sec:broadKnown}. The internally calibrated RP spectrum for one example, {\tt Gaia DR3 2067888218857234304}, with $G \approx 9.246^{\rm m}$ and $G_{BP} - G_{RP} \approx 2.008^{\rm m}$, is shown in Fig.~\ref{fig:dib}. \cite{Maiz2021} report an equivalent width of the 770nm band of $(-1.49 \pm 0.20)$nm for this star. An extremum in the second derivative, consistent with the pseudo-wavelength position of the 770nm band, is detected, with a high $p-$value of 0.9996. The continuum model is shown as a red line in Fig.~\ref{fig:dib}, derived for a Gaussian band profile. The resulting equivalent width of the band of $W = (-1.38 \pm 0.55)$nm is well in agreement with the result by \cite{Maiz2021}. The line width parameter $\omega$ however is only $1.15 \pm 0.16$, and the band is thus still in agreement with being narrow. It is nevertheless consistent with the model for the pseudo-LSF used, which has a FWMH of already 22.7nm at a wavelength of 770nm. The expected value of $\omega$ is 1.21, slightly larger than unity only, and also in good agreement with the measured value of $\omega$. The line parameters for the 770nm band are listed in Table~\ref{tab:WD}, too.

\section{Summary and discussion \label{sec:summary}}

We derived a methodology for the analysis of spectral emission and absorption lines in \gaia~DR3 low resolution spectra. These spectra are represented by linear combinations of Hermite functions in their form of internally calibrated spectra, and we exploit the useful mathematical properties that come with this form of representation. These properties include that the derivative of a linear combination of Hermite functions is another linear combination of Hermite functions, that can be derived by a matrix-vector multiplication. Furthermore, all roots of a linear combination of Hermite functions can be computed as eigenvalues of a simple non-standard companion matrix. These properties combined allow for a convenient and efficient determination of local extrema and inflection points in \gaia~DR3 XP spectra. We used this to derive expression for the position, error on position, statistical significance, and equivalent widths of spectral lines, both for narrow and broad lines.\par
The resulting computations are very fast and robust and therefore can be applied to search for spectral lines and extract their parameters even for large numbers of objects with modest computational resources. Neither machine learning nor human interaction is required.\par
We applied the derived methods to \gaia~DR3 low resolution spectra with hydrogen Balmer lines, He I lines, and a broad interstellar band. The line positions, widths, and equivalent widths, as well as the statistical significance of the corresponding local extrema in the spectra were computed as a demonstration. The presented methodology based on the derivatives of Hermite functions might help to make use of the interesting and in its form unique new \gaia~data product that are the low resolution spectra. Despite their low spectral resolution, \gaia~XP spectra provide opportunities for the analysis of absorption and emission lines from the near-ultraviolet to the near-infrared part of the electromagnetic spectrum. Exploiting the mathematical properties of the unique representation of the spectra as linear combinations of Hermite functions allows for a detailed analysis of the spectral lines.

\begin{acknowledgements}
This work was funded by the Spanish MICIN/AEI/10.13039/501100011033 and by ''ERDF A way of making Europe'' by the European Union through grant RTI2018-095076-B-C21, and the Institute of Cosmos Sciences University of Barcelona (ICCUB, Unidad de Excelencia 'Mar\'{i}a de Maeztu') through grant CEX2019-000918-M.\\
      This work has made use of data from the European Space Agency (ESA) mission {\it Gaia} (\url{https://www.cosmos.esa.int/gaia}), processed by the {\it Gaia} Data Processing and Analysis Consortium (DPAC, \url{https://www.cosmos.esa.int/web/gaia/dpac/consortium}). Funding for the DPAC has been provided by national institutions, in particular the institutions participating in the {\it Gaia} Multilateral Agreement.
\end{acknowledgements}

\bibliographystyle{aa} 
\bibliography{SpectralLines} 

\end{document}